\documentclass[times,12pt,twocolumn]{aastex631}
\usepackage{graphicx}
\graphicspath{ {images/} }
\usepackage{wrapfig}
\usepackage[rightcaption]{sidecap}
\usepackage{xspace}
\usepackage{gensymb}
\usepackage{amsmath}
\usepackage{bm}
\usepackage{wasysym}

\newcommand{\Ha}	{H$\alpha$}%
\newcommand{\Hb}	{H$\beta$}%
\newcommand{\BD}	{H$\alpha$/H$\beta$}%
\newcommand{\rev}   {}

\begin{document}

\title{Ground- and Space-Based Dust Observations of VV\,191 Overlapping Galaxy Pair}

\author[0000-0002-5404-1372]{Clayton Robertson} 
\affiliation{Department of Physics, University of Louisville, Natural Science Building 102, Louisville KY 40292, USA}

\author[0000-0002-4884-6756]{Benne W. Holwerda}
\affiliation{Department of Physics, University of Louisville, Natural Science Building 102, Louisville KY 40292, USA}

\author[0000-0002-5830-9233]{Jason Young}
\affiliation{Department of Astronomy, University of Massachusetts, Amherst, MA 01003, USA}

\author[0000-0002-6131-9539]{William C. Keel}
\affiliation{Department of Physics and Astronomy, University of Alabama, Box 870324, Tuscaloosa, AL 35404, USA}


\author[0000-0001-6265-0541]{Jessica M. Berkheimer}
\affiliation{School of Earth \& Space Exploration, Arizona State University, Tempe, AZ\,85287-1404, USA}

\author[0000-0002-4012-779X]{Kyle Cook}
\affiliation{Department of Physics, University of Louisville, Natural Science Building 102, Louisville KY 40292, USA}

\author[0000-0003-1949-7638]{Christopher J. Conselice}
\affiliation{Jodrell Bank Centre for Astrophysics, University of Manchester, Oxford Road, Manchester M13 9PL, UK}

\author[0000-0003-1625-8009]{Brenda L.~Frye}
\affiliation{Department of Astronomy/Steward Observatory, University of Arizona, 933 N. Cherry Avenue, Tucson, AZ 85721, USA}

\author[0000-0001-9440-8872]{Norman A.~Grogin} 
\affiliation{Space Telescope Science Institute, 3700 San Martin Drive, Baltimore, MD 21218, USA}

\author[0000-0002-6610-2048]{Anton M. Koekemoer} 
\affiliation{Space Telescope Science Institute, 3700 San Martin Drive, Baltimore, MD 21218, USA}

\author[0000-0001-9365-369X]{Camella Nasr}
\affiliation{Orion Space Solutions, Louisville, CO 80027}

\author[0009-0007-2620-7033]{Divya Patel}
\affiliation{Department of Physics, University of Louisville, Natural Science Building 102, Louisville KY 40292, USA}

\author[0009-0008-4152-5088]{Wade Roemer}
\affiliation{NASA Marshall Space Flight Center ES52}

\author[0000-0002-8280-5798]{Dominic Smith}
\affiliation{Department of Physics, University of Louisville}

\author[0000-0001-8156-6281]{Rogier A. Windhorst} 
\affiliation{School of Earth and Space Exploration, Arizona State University,
Tempe, AZ 85287-1404, USA}



\begin{abstract}
The Balmer decrement (\BD) provides a constraint on attenuation, the cumulative effects of dust grains in the ISM. The ratio is a reliable spectroscopic tool for deriving the dust properties of galaxies that determine many different quantities such as star formation rate, metallicity, and SED models. Here we measure independently both the attenuation and \BD\ of an occulting galaxy pair: VV\,191. Attenuation measurements in the visible spectrum (A$_{V,stars}$) from dust maps derived from the F606W filter of HST and the F090W filter of JWST are matched with spaxel-by-spaxel \BD\ observations from the George and Cynthia Mitchell Spectrograph (GCMS) of the McDonald Observatory. The 0.5 to 0.7 micron bandpass covers the Balmer lines for VV\,191. The dust maps of JWST and HST provide the high sensitivity necessary for comparisons and tracking trends of the geometrically favorable galaxy. We present maps and plots of the Balmer lines for the VV\,191 galaxy pair and for a specific region highlighting dust lanes for VV\,191b in the overlap region. \rev{We compute A$_{V, HII}$ from \BD\ and plot both quanities against A$_{V, stars}$. Our results show that regions with higher dust content, residing closer to the spiral center, dominate ionized gas attenuation, leading to an overestimation of A$_{V, HII}$ by a factor or 2. Further out in the spiral arms, the lower dust content leads to more agreement between the attenuations, indicating lower SFR and larger contribution from older stars to the stellar continuum outside the Petrosian radius.} 
\end{abstract}

\section{Introduction}

Dust is a key factor in all astronomical observations of the interstellar medium (ISM). It plays a role in how a galaxy is physically observed and heavily influences astronomical distance measurements. To provide an accurate analysis of galaxies in the ISM, knowledge of the cumulative effect of how dust grains are traced in the galaxy is imperative. This is accomplished by attenuation: the combination of extinction and scattering. Dust attenuation remains a fundamental uncertainty in extragalactic astronomy while profoundly impacting the derivation of the physical properties of galaxies in the ISM. \rev{Differences among attenuation laws greatly influence the measurement and calculation of physical properties such as star formation rate (SFR), stellar mass, and overall dust content \citep{salim_dust_2020, kriek_dust_2013, shivaei_investigating_2015}.} All stellar population and Spectral Energy Distribution (SED) models of galaxies -- and many distance measurements (e.g. SNIa) -- assume the dependence of the attenuation on wavelength. This is commonly referred to as the attenuation curve or extinction law. The relation used is typically linked to the Balmer decrement \citep[see]{conroy_dust_2010,wild_empirical_2011,kriek_dust_2013,kreckel_mapping_2013,reddy_mosdef_2015,reddy_spectroscopic_2016} or as a variable in SED fits \citep[e.g.][]{faro_characterizing_2017,tress_shards_2018}.

\rev{Different attenuation relations between sightlines in the Magellanic clouds \citep{gordon_starburstlike_1998,misselt_reanalysis_1999,li_infrared_2002,hagen_evolution_2015} and our Milky Way \citep{cardelli_relationship_1989,fitzpatrick_correcting_1999} demonstrate that there is a wide variety in attenuation relations for the local ISM. Explanations of this observed diversity may involve geometry between the dusty ISM and stars, metallicity, or sampling biases created by emission lines. \citep{calzetti_dust_1994,calzetti_dust_2000,gordon_quantitative_2003,charlot_simple_2000,reddy_mosdef_2015,wild_empirical_2011}} 

It would be preferable to have a method of measuring both the attenuation curve and the Balmer decrement independently. Here, we explore this with James Webb Space Telescope (JWST) NIRCam, Hubble Space Telescope (HST) Wide Field Camera 3, and the George and Cynthia Mitchell (GCMS; formerly known as VIRUS-P) ground-based IFU observations of a nearby occulting galaxy pair -- a coincidental overlap of two galaxies -- assuming symmetry around their center. Information about the dust structure of the foreground galaxy is found by analyzing the absorption line features produced in the spectrum of the background galaxy. \cite{white_direct_1992} pioneered the estimation of dust attenuation from differential photometry in occulting galaxies. The technique was first applied to ground-based observations \citep{andredakis_photometry_1992, berlind_extinction_1997,domingue_dust_1999,white_iii_seeing_2000} and then to space-based imaging \citep{elmegreen_italhubble_2001,holwerda_vltvimos_2013} and spectroscopy \citep{holwerda_occulting_2013,holwerda_vltvimos_2013}.



The GCMS spectral range covers the first two lines of the Balmer series  of the foreground galaxy. Since the intrinsic ratio of these lines are nearly universal, the relative intensity of these two strong emission lines is a common extinction estimate of star-forming regions. The map of the Balmer decrement (\BD) and the attenuation slope can directly be compared here to those from models \citep[e.g.,][]{charlot_simple_2000, wild_empirical_2011}. A pressing question emerges as to whether the Balmer decrement accurately predicts the attenuation law.

We present the combination of JWST/HST dust attenuation maps and spectroscopic observations from GCMS to compare two independently derived quantities: the Balmer decrement and dust attenuation in the visible spectrum \rev{derived from the method of occulting galaxy pairs \citep[see][]{white_direct_1992,holwerda_occulting_2013,keel_jwsts_2023}}, A$_{V, stars}$. We aim to juxtapose A$_{V}$ to characteristics in the foreground disk traced by the {\sc HII} emission lines using a galaxy pair with nearly ideal geometry for this analysis.  


This paper is organized as follows: 
Section \ref{s:vv191} describes our target, the overlapping pair VV\,191. Section \ref{s:data} describes the JWST/HST data, the GCMS IFU observations and data reduction.
Section \ref{s:map} describes how the attenuation and Balmer decrement are mapped in the foreground galaxy. Section \ref{s:BalmerLit} covers the use of the Balmer decrement to determine the dust attenuation in galaxies in the literature.
Section \ref{s:results} presents our results.
Sections \ref{s:discussion} and \ref{s:conclusions} comprise our discussion and conclusion, respectively. 
\\


\section{VV191 Galaxy Pair and its Properties}
\label{s:vv191}

There are over 2000 known occulting galaxy pairs in the local universe, thanks to the efforts of Galaxy Zoo \citep{lintott_galaxy_2008,keel_galaxy_2013}. \rev{Further analysis by STARSMOG\footnote{STarlight Attenuation and Reddening in a Survey of Multiple Overlapping Galaxies} of Hubble Space Telescope snapshots resulted in an additional 50 pairs.} STARSMOG determined that VV\,191 has the ideal configuration: a partially overlapping, bona fide occulting pair, i.e., overlapping on the sky but physically well-separated in distance. 

According to the VV atlas \citep{vorontsov-velyaminov_atlas_1959}, the VV\,191 galaxy pair consists of VV\,191a, the elliptical background galaxy, and VV\,191b, the spiral galaxy located northeast to VV\,191a. Both galaxies are large in the sky. \cite{keel_jwsts_2023} determined the Petrosian radius of the spiral foreground as 12.5" = 12 kpc in the F606W filter. The background elliptical galaxy extends to 1.7 Petrosian radii away from the foreground spiral's center, backlighting dust roughly 20 kpc in radius.

\subsection{Possible Gravitational Interaction?}
\label{s:grav interact}

The difference in redshift between the two galaxies in VV\,191 is too small ($\Delta z = 0.001$) to completely rule out the possibility of gravitational interaction in the pair's past or present. \rev{The stellar and dust structures of physically close pairs are} typically distorted enough due to tidal interactions that they are of limited use \rev{to study dust attenuation from overlapping galaxy pairs.} Tidal interaction is a point of interest in occulting galaxy pairs due to the possibility of compromising the symmetry assumed for each galaxy \citep{holwerda_occulting_2013}, especially for the spiral foreground. Early on, the significant redshift difference made VV\,191 an ideal candidate. A large enough redshift difference indicated little to no mutual gravitational interaction between the galaxies. However, the updated redshifts do not fully answer the question of interactions. Thus, any evidence of asymmetry that suggests gravitational interaction \rev{requires} visual inspection with high-resolution HST and JWST images. From visual analysis, neither galaxy shows any asymmetry that suggests interaction within their Petrosian radii of 12 kpc \citep{keel_jwsts_2023}.



\section{Data} 
\label{s:data}

Data for this project came from three observatories: JWST (full data reduction and analysis by \cite{keel_jwsts_2023}, HST, and ground-based GCMS IFU observations at McDonald Observatory.

\subsection{JWST and HST Data}

VV\,191 was selected as a target for the Guaranteed Time Program ``Prime Extragalactic Areas for Reionization and Lensing Science'' (PEARLS, PI: R.~Windhorst; \cite{windhorst_jwst_2023}) based on its identification as an ideally configured pair by an HST snapshot program. 

\subsubsection{HST WFC3}

An initial HST Wide Field Camera 3 image of VV\,191 was obtained as part of the STARSMOG snapshot program (HST program 13695, PI B.W. Holwerda), which targeted potential backlit-galaxy systems from the Galaxy Zoo \citep{keel_galaxy_2013} and Galaxy and Mass Assembly (GAMA) survey \citep{holwerda_galaxy_2015} samples. Snapshots are in the F606W filter with a total exposure time of 900 seconds in two sub-exposures with a small dither motion between. 

In preparation for the JWST ERS program, a second HST/WFC3 observation was \rev{carried out} in the UV F336W and F225W bands for 2749 and 8874 seconds, respectively (HST program 15106, PI B. W. Holwerda). Lower flux levels and cosmic ray events made attenuation maps in these filters a challenge \citep[see][]{keel_jwsts_2023}.

\subsubsection{JWST}

VV\,191 was observed with the JWST \citep[JWST,][]{menzel_design_2023,mcelwain_james_2023,gardner_james_2023} using NIRCam \citep{rieke_performance_2023} on 2 July 2022 as part of the PEARLS GTO program (1176).
The VV\,191 pair fits in a single short-wavelength detector quadrant, centered on the dust-overlap region, allowing for freedom of orientation, ease of scheduling, and simpler reduction. Exposure times were 901 seconds in each of the filters (F090W, F150W, F356W and F444W) using MEDIUM8 readout as a three-point dither strategy. The JWST/NIRCam data is presented in full in \cite{keel_jwsts_2023}, including the steps to construct attenuation maps (see below).

\subsection{Ground-based IFU Observations From GCMS} 

To supplement the JWST/HST observations with IFU data, we observed this pair with GCMS at the McDonald Observatory in Spring 2022.

GCMS is hosted at the 2.7 m Harlan J. Smith telescope at the McDonald Observatory \citep{tufts_virus-p_2008, hill_design_2008}. The IFU is comprised of 246 diameter fibers that are distributed in a fixed hexagonal pattern with a filling factor of roughly one-third. At the distance of VV\,191, the fiber diameter corresponds to 1.4 kpc sampling and the field of view is $\simeq$ 35 kpc per side.

The observations were made with the VP1 grating and covered the red 4500 - 7000 \AA\ spectral range with a spectral resolution of 5.3 \AA. 
We opted for the modified red configuration, with an adjusted spectral coverage (4500 - 7000 \AA\ instead of the typical 4350 - 6850 \AA) in order to include both $H\alpha$ and $H\beta$ in a single observation. This serves the principal science goal of constraining the Balmer decrement.

Our aim is to \rev{maximize signal from the} overlapping parts of both galaxies to reach as low surface brightness as practical for both emission lines. Nominal limits are the extent of the disk (R25), which is 25.5 mag/arcsec$^2$. Aiming for a S/N $\sim3$ at this outer limit of the foreground spiral galaxy, and a dither strategy, we estimated at 5076 sec per pointing for a total of 8.46 hours to obtain sufficient signal-to-noise \rev{(S/N)}. 

To ensure a high filling factor, a six-pointing dither pattern was followed, resulting in a 738-fiber coverage of the field of view. \rev{The dithering pattern was designed to achieve a 100\% fill factor and maximize the spatial sampling.} 



Observations were carried out on May 24-25 and 25-26 of 2022. Sky conditions were optimal on the second night. Each six-point dither pattern was repeated 3x with either 700 or 800s exposures at each pointing. A guider issue interrupted the fifth iteration of the six-point dither, but the observations were otherwise uneventful. Total exposure time on target was 4500s. Throughout the acquisition, the full width at half maximum (FWHM) of the seeing varied between 1.5 and 2 arcseconds. Standard stars were observed both nights.  \\

\subsubsection{GCMS Data Reduction}
The GCMS data reduction follows the pipeline used in \cite{young_distribution_2015,young_recent_2020}: basic data-reduction (bias, flat-field through dome flats, and wavelength calibration through arc line exposures), sky subtraction, reconstruction of the field-of-view reconstruction and spectrophotometric calibration.

Wavelength calibration is done with dedicated arc lamp exposures at the beginning of each night. This results in a wavelength calibration better than the spectal sampling (2 \AA). 
Sky lines, e.g. the one at 5757 \AA, can be identified in the original raw data but the final cube is the combination of many exposures, averaging out the residual. This does leave a feature at $\sim$ 5300 \AA (Figure \ref{fig: spectra}) which behaves as a sky line but may be reflected artificial light. 

\cite{young_distribution_2015} developed a GCMS reduction pipeline optimized for low surface brightness, ideal for our purposes. Once reduced (bias, flatfielded and spectral calibration), the GCMS spectra were mapped onto a spectral data cube with a 0\farcs5 plate scale and a 2 \AA\ spectral resolution. Flux units for the spectra are in $erg s^{-1} cm^{-2}$ \AA$^{-1}$ arcsec$^{-2}$ in the following figures. 

The angular resolution in our spectral data cubes is limited partly by the seeing at Mt. Locke and the 4\farcs16 diameter of the GCMS fibers. The  6-point dither pattern observing strategy allows a reconstruction of the image plane from the fiber data to an angular resolution of around 2$^{\prime\prime}$. We have chosen to reconstruct the image plane at a 0\farcs5 spatial sampling to properly sample this angular resolution. VV\,191 fits comfortably in the dithered FOV of GCMS.

\section{Mapping the Attenuation Curve from Occulting Galaxy Pairs}
\label{s:map}

\subsection{Dust in Occulting Galaxies} 


To map dust extinction, a known background source of light is required. In the case of a spiral galaxy overlapping a more distant galaxy, the most distant galaxy is the known light source, assuming it has a symmetric light distribution. Originally proposed by \cite{white_direct_1992}, the sample size initially limited this technique, and all known pairs were quickly analyzed. 
Fortunately, more pairs were found in SDSS \citep{holwerda_spiral_2007}, the Galaxy And Mass Assembly (GAMA) spectra \citep{holwerda_galaxy_2015}, and by Galaxy Zoo, totaling approximately 2500 pairs.

Analysis in \cite{holwerda_extended_2009} showed an overlapping pair within HST's Advanced Camera for Surveys (ACS) observations of NGC 253. The average reddening-attenuation slope closely resembled the Milky Way extinction law: R$_V$ = 3.1. However, the relation between A$_V$ and the color excess, $E (B - V) = A_B - A_V$ where $A_B$ is the attenuation in the blue, showed significant scatter, which invites further questions and examinations of spiral disks in the ISM.

An early integral field unit (IFU) study of galaxy pair 2MASX J00482185 - 2507365 with VIMOS \citep[][]{holwerda_occulting_2013} showed the benefits and limitations of small-FOV, coarse IFU observations. This almost optimal pair showed an extended dusty disk around the foreground galaxy, extending well beyond its optical size. This disk's attenuation curves show moderate attenuation and a range of slopes ($A_V$ and $R_V$, respectively), but this is limited by spatial sampling. 



\subsection{Mapping the Attenuation-Reddening Curve} 


\cite{keel_seeing_2001} and \cite{holwerda_extended_2009} found from photometry that the {\em mean} attenuation curve becomes galactic when the physical sampling scale is under $\sim$100 pc. In many galaxies, there is a high and low attenuation component, and HST can discriminate their ratio within a GCMS element, eliminating the degeneracy in the covering factor (the ``Matryoshka" Effect). Hence, to sample the {\em variance} in attenuation curves, the IFU spatial resolution needs to be similar or better, having a distance limit of $z<0.05$ for the foreground galaxy.

The principal uncertainty in the overlapping galaxy method is the assumption of rotational symmetry of both galaxies. In the case of IFU observations, as we cross-correlate the spectra from corresponding pixels, we can relax some symmetry requirements \citep[see][for discussion on single-slit observations]{domingue_seeing_2000}. 

The observational requirements of overlapping galaxies detected with an IFU are atypical, as these put a premium on field-of-view, spatial sampling, and spectral coverage while having no spectral resolution requirements. These requirements prefer a face-on galaxy pair: the image would give a proper field-of-view for detecting possible factors such as rotational asymmetries in the foreground spiral disk, while displaying the backlit overlapping section with its complementary non-overlapping components. GCMS is an excellent fit for such observations as it has exquisite sensitivity and enough spatial resolution to sample our target galaxy pair while having enough spectral coverage to encapsulate the {\sc H ii} regions of the foreground galaxy, thereby allowing us to map \BD\ for each IFU spaxel captured by the instrument. 

The goal is to compare the attenuation curve (difference in continuum) to characteristics in the foreground disk traced by emission lines. GCMS's combination of FOV (1\farcm7), spatial sampling (4\farcs16 oversampling with a 6-point pattern), and spectral range allow us to properly observe galaxy overlaps and sample the foreground galaxy's ISM with sufficient detail.

\subsection{Spectra}

\begin{figure*}
    \begin{minipage}{\textwidth}
    \centering
        \includegraphics[width=\textwidth]{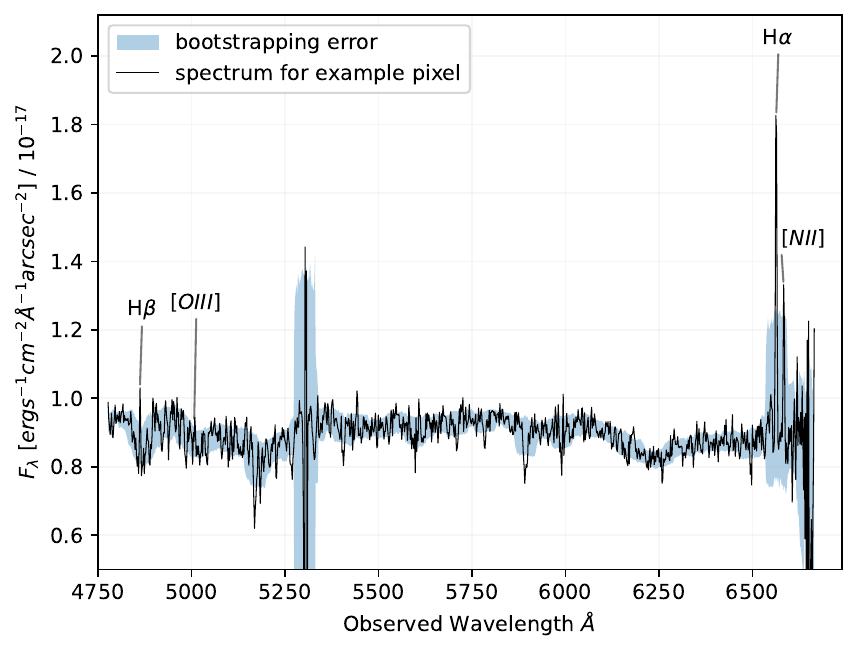} 
    \end{minipage}
    \caption{The redshifted spectrum for an example spaxel from GCMS. This spaxel is located in the overlap region nearby the center, approximately 5.63 kpc from the galaxy center. The location of the pixel is Right Ascension (RA): 207.0931702$^{\degree}$, Declination (Dec): 25.6792436$^{\degree}$. 1 spaxel is approximately 0.001389$^{\degree}$ for both RA and Dec. The Balmer emission lines (\Ha\ and \Hb) along with the other sky lines within the available bandwidth are also shown. Wavelengths are multiplied by 1.0513 to account for the redshift of VV\,191b, where z = 0.0513. The known sky lines at 5577\AA{} is successfully removed and mitigated in the combination of individual exposures into a single cube. The feature at $\sim$5300\AA{} is suspected to be reflected artificial light, and the absorption feature at $\sim$ 5900\AA{} is a known flat-fielding effect which is telescope position dependent. Both \rev{of} these features will be corrected in a future iteration of the pipeline. 
   }
    \label{fig: spectra}
\end{figure*}

\begin{figure}
    \begin{minipage}{0.49\textwidth}
        \includegraphics[width=\textwidth]{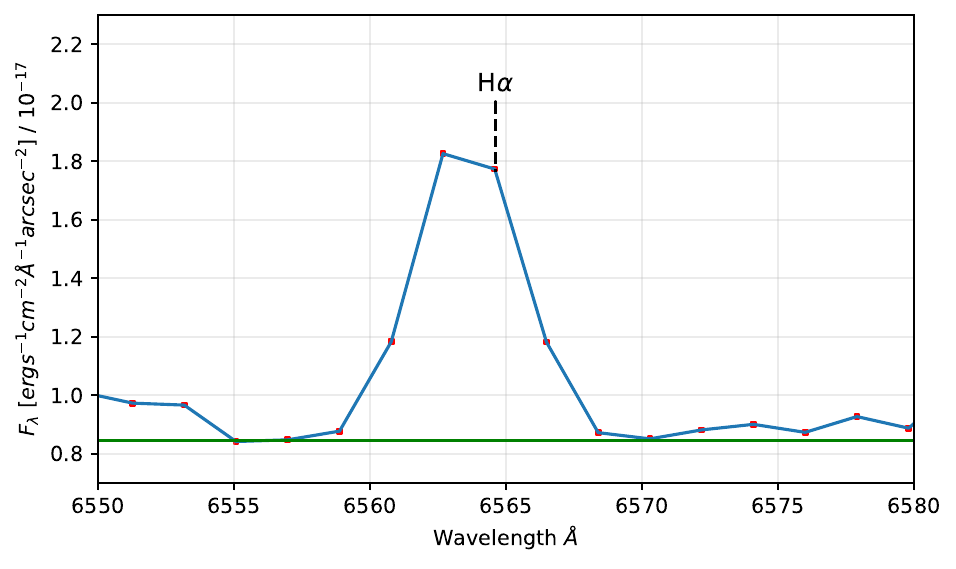}
    \end{minipage}
    \begin{minipage}{0.49\textwidth}
        \includegraphics[width=\textwidth]{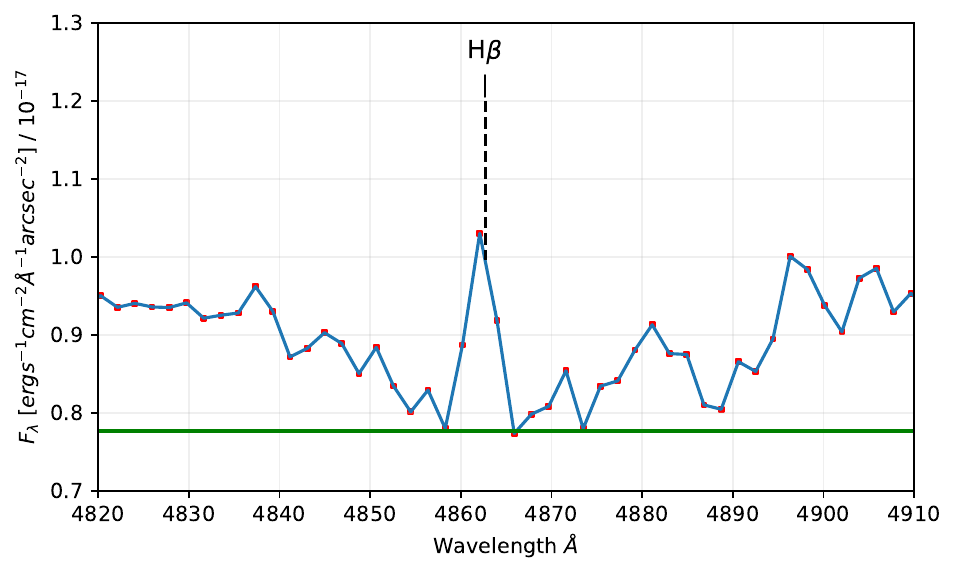}
    \end{minipage}
    \caption{\Ha\ and \Hb\ peaks peaks zoomed in from same example spaxel taken by GCMS as in Figure \ref{fig: spectra}. The continuum (horizontal green line) is derived from interpolated data for both H$\alpha$ and H$\beta$, calculated by taking the sigma clipped mean of the outermost pixels of the peak along with two pixels further out. The red pixels represent each pixel in the spectra (same units) taken for the spectra. The units for the flux $F_{\lambda}$ are the same as defined in Figure \ref{fig: spectra} and divided by a normalization factor of $10^{-17}$. Spectrum is plotted with the \textsc{line id plot} package.}
    \label{fig:spectra zoom}
\end{figure}

The entire sample spectrum extracted \rev{from a sample spaxel from the} available GCMS IFU data is shown in Figure \ref{fig: spectra}. \rev{The error spectrum is shown in blue. The error for each pixel is estimated from the mean and standard deviation from a 30 pixel window in the spectrum centered at that said pixel. }

The Python module Line Identification Plots With Automatic Label Layout (Copyright (c) 2011, Prasanth Nair) marks the Balmer lines (Figure \ref{fig:spectra zoom}) examined in this paper. There are some interesting notes to point out from these spectra: 
\begin{enumerate}
\item The spectra are flat due to limited bandwidth (close to 2000 \AA). 
\item For nearly every single spaxel spanning the entire galaxy pair, an unusual phenomenon of large absorption/emission occurs around 5300 \AA. \rev{The origin of this spike is unclear, though we suspect it may be an artifact of the instrumentation or artificial light.} It does not impact our emission line measurements. 
\item The \textsc{[OIII]} and {\sc [NII]} emission lines \rev{are labeled with the} Balmer lines in Figure \ref{fig: spectra}. They are considerably weaker than the Balmer lines. \rev{We use the same method to extract \textsc{[OIII]} and {\sc [NII]} as done with the Balmer lines to attempt to constrain the metallcity via the R23 \citep{tremonti_origin_2004} and the N2 \citep{pettini_o_2004, denicolo_new_2002} line ratios \citep{kewley_metallicity_2008}. We choose to omit the results as they are not reliable due to these lines being not formally detected. We hope to revisit the metallicity for the VV\,191 with future observations.}


\end{enumerate}

\subsection{Extracting the Emission Lines}

The Balmer Decrement is obtained from the flux from the H$\alpha$ and H$\beta$ emissions lines from each spaxel generated from a 3-D cube FITS file that encompasses the entire VV\,191 galaxy pair. Each spaxel of the GCMS data has its own spectrum similar to Figure \ref{fig: spectra}. The H$\alpha$ and the H$\beta$ peaks are labeled. 

\rev{Complications of the continuum fittings arise from the surrounding photospheric absorption from O, B, and A stars \citep{delgado_stellar_2000}.} Absorption is most prevalent with \Hb\, as shown in Figure \ref{fig: spectra}, which brings the value of the continuum below the stellar continuum for the entire spectra. Initially, linear interpolation was used on the peak edges to account for the stellar continuum. However, uneven and elevated interpolated values due to noisy spaxels outside of the Petrosian radius \rev{led to an underestimation of} \Hb\ emission lines. The continuum was measured by taking the average of the two pixels on the outside of the peak edges. The emission peaks were calculated with Simpson integration, from which the emission flux, continuum, and equivalent widths could be immediately calculated. \rev{The uncertainties were estimated in our emission line measurements by \citep{tresse_spectral_1999}}

\rev{\begin{equation}
    \sigma_{F} = \sigma_{c} D \sqrt{2N_{pix} + EW/D}
    \label{flux_uncertainty}
\end{equation}}

\rev{where $\sigma{c}$ , is the mean standard deviation per pixel of the continuum, D is the spectral dispersion of per pixel in \AA\, $N_{pix}$ is the number of pixels covered by each line, and EW is the equivalent width. $\sigma_{c}$ is estimated from the from the two values on each side of the emission line.}


\subsection{Balmer Decrement} 
\label{s:Balmer Dec}

Practically, the Balmer Decrement has always been effective in the determination of dust extinction and attenuation dating back to the work of \cite{kennicutt_integrated_1992} and has been improved upon ever since \citep[e.g.,][]{calzetti_dust_2001, dominguez_dust_2013, groves_balmer_2012}. The Balmer lines are well defined from quantum physics; the electron in the hydrogen atom transitioning from the n = 3 level to the n = 2 level is the H$\alpha$ emission peak, showing up around 6564 \AA\ (essentially red). The electron transition from n = 4 to the n = 2 level is the H$\beta$ emission peak. Since it releases more energy than H$\alpha$, it shows up bluer (around 4862 \AA). The intrinsic value for the Balmer decrement under unattenuated conditions varies depending on the type of galaxy observed. A value of 2.86 is assumed for galaxies dominated by star formation, while a value of 3.1 is assumed for Active Galactic Nuclei (AGN)-dominated galaxies \citep{osterbrock_astrophysics_2006}. We select the former value of $(H\alpha/H\beta)_{int}$ = 2.86 for VV\,191b because it does not have an AGN. We set the {\sc HII} regions with a Case B recombination at 10,000 K and electron density of around 10$^2$ cm$^{-3}$. The difference between the observed value and the intrinsic value for the Balmer decrement is attributed to dust attenuation under specific conditions.

\section{Relation between Balmer Decrement and Dust Attenuation in the Literature}
\label{s:BalmerLit}

We will observe how the attenuation is dependent on the Balmer decrement by comparing \rev{the Balmer line ratio to commonly used relations that utilize both \BD\ and A$_{V}$.} The works of \cite{calzetti_dust_2001} and \cite{salim_dust_2020} are \rev{two detailed reviews of dust extinction and attenuation laws in galaxies.} To show how it is reliant on \BD\ we follow the derivation from the Appendix in \cite{momcheva_nebular_2013}, which utilizes the Balmer color excess. The relation between the optical depth, $\tau(\lambda)$, the intrinsic source of intensity, $I(\lambda)_{int}$, and the observed intensity, $I(\lambda)_{obs}$, is given as:

\begin{equation}
    e^{-\tau(\lambda)}=\frac{I(\lambda)_{obs}}{I(\lambda)_{int}},
    \label{eq1}
\end{equation}

The ratio of the intensities on the right-hand side of Equation \ref{eq1} is also representative of the transmission, $\overline{T}$, which is related to the optical depth and attenuation by:

\begin{equation}
    A_V = 1.086 \times \tau(\lambda) = -1.086 \times \ln{\overline{T}},
    \label{eq transmission}
\end{equation}

where the intensities and the optical depth depend on the wavelength $\lambda$.  We define the Balmer color excess as:

\begin{equation}
    E(H\beta - H\alpha) =  2.5\log{ \left (\frac{H\alpha/H\beta}{(H\alpha/H\beta)_{int}} \right)}.
    \label{eq balmer excess}
\end{equation}

The broadband color excess, $E(B - V)$, is related to Equation \ref{eq balmer excess} by:

\begin{equation}
    E(H\beta - H\alpha) = E(B - V)(k_{H\beta} - k_{H\alpha})
    \label{eq4}
\end{equation}

This requires the adoption of $k_{\lambda}$: the total-to-selective extinction measured at the wavelength of the emission line. The total attenuation, A$_{\lambda}$ is directly related to the broadband color excess by \citep{reddy_mosdef_2015}:

\begin{equation}
    A_{\lambda} = k_{\lambda} \times E(B-V).
    \label{eq color excess}
\end{equation}

We combine Equations \ref{eq balmer excess}, \ref{eq4}, and \ref{eq color excess} to get the total attenuation in the visible spectrum, A$_{V}$:

\begin{equation}
\begin{split}
    A_{V} & = k_{V} \times E(B-V) \\
          & = 2.5\log{ \left (\frac{H\alpha/H\beta}{(H\alpha/H\beta)_{int}} \right)} \times \frac{k_{V}}{k_{H\beta} - k_{H\alpha}}.
\end{split}
\label{eq av and bd}
\end{equation}

For wavelengths in the visible spectrum, k$_{V}$ approaches R$_{V}$. From \cite{calzetti_dust_2000}:

\begin{equation} 
    k_{\lambda}=\left\{
    \begin{array}{lr}
       2.659(-2.156 + 1.509/\lambda \\
       - .198/\lambda^2 +.011/\lambda^3)\\
       + R_{V}  &  .12 \mu m \leq \lambda \leq .63 \mu m\\
      \\
       2.659(-1.857 + 1.040/\lambda)\\
       + R_{V}  & .63 \mu m \leq \lambda \leq 2.2 \mu m\\
    \end{array}
    \right\}
\label{eq:k values} 
\end{equation}

We are interested when $\lambda$ = 0.606 $\mu m$, or 6060 \AA, as the attenuation derived from HST/JWST is from the F606W filter, which corresponds to the specified wavelength. Applying the equation above gives k$_{V}^{Calzetti00}$ = R$_{V} - 0.41$. \rev{\cite{reddy_mosdef_2015} gives a similar equation relating $k_{V}$ to $R_{V}$ (their Equation 7) and gives a similar result k$_{V}^{Reddy15}$ = R$_{V} - 0.37$. Given how close the values for $k_{V}$ are, we do not find any significant differences using one relation over the other. So, for simplicity we assume the \cite{calzetti_dust_2000} law for the remainder of this paper, i.e. $k_{V}^{Calzetti00} = k_{V}$. This correction will be applied to Equation \ref{eq av and bd}, as most varying relations differentiate based on the R$_{V}$ values. The values of $k_{H\alpha}$ and $k_{H\beta}$ are calculated from Equation \ref{eq:k values} as well.  }

\section{Results}
\label{s:results}

\subsection{Redshifts and Possible Incline of VV 191b} 

The easily identifiable \Ha\ emission lines measure the spectroscopic redshifts by comparing the observed emission lines with their theoretical vacuum wavelength. This is done almost automatically with the Python package {\sc LiMe} (https://lime-stable.readthedocs.io). 

First, to measure the inclination of the face-on spiral galaxy, windows on each side, left and right, equidistant from the center, are taken from the FITS cube. Each window is a $20 \times 20$ box to create 400 redshift values for each side. The center spaxel used for calculating the redshift on the west side, z$_{W}$, has an RA and Dec of 207.095$^{\circ}$ and 25.680$^{\circ}$ respectively. The redshift on the east side, z$_{E}$, has its center spaxel location of RA = 207.091$^{\circ}$ and Dec = 25.680$^{\circ}$. Both windows \rev{have very minimal standard deviation with mean redshifts of} $z_{W} = 0.05135\, \pm\, 1.7\times10^{-17}$ and $z_{E} = 0.05156\, \pm\, 9.4\times10^{-6}$. The small error confirms the slight inclination of $\sim$ 1200 km/s, with the east side receding and the west side approaching.

Next, we compare the overall redshift of the center of VV\,191b to the redshift logged by SDSS to determine any differences. We executed the same method of utilizing {\sc LiMe}, instead using the center of VV\,191 and not the east/west portions of the spiral galaxy. The center window is larger ($10 \times 10$ spaxels) to create a larger sample size of 100 redshift values. We find that the mean redshift is z$_{center}$ = 0.05137 with little deviation ($\sigma \simeq 10^{-5}$). Updated redshifts from \cite{keel_jwsts_2023} show that z = 0.0513 and z = 0.0514 for VV\,191a and VV\,191b, respectively. The differences in redshift from SDSS, \cite{keel_jwsts_2023}, and this paper are minimal, but noted and considered when determining the center peaks of \Ha\ and \Hb. To avoid confusion, we will use results from \cite{keel_jwsts_2023} of the redshift for the remainder of the paper, as those values \rev{are based on the best available spectroscopy.}

\subsection{Balmer Decrement Maps}

\begin{figure*}
    \begin{minipage}{0.49\linewidth}  
        \includegraphics[width=\textwidth, height=\textwidth]{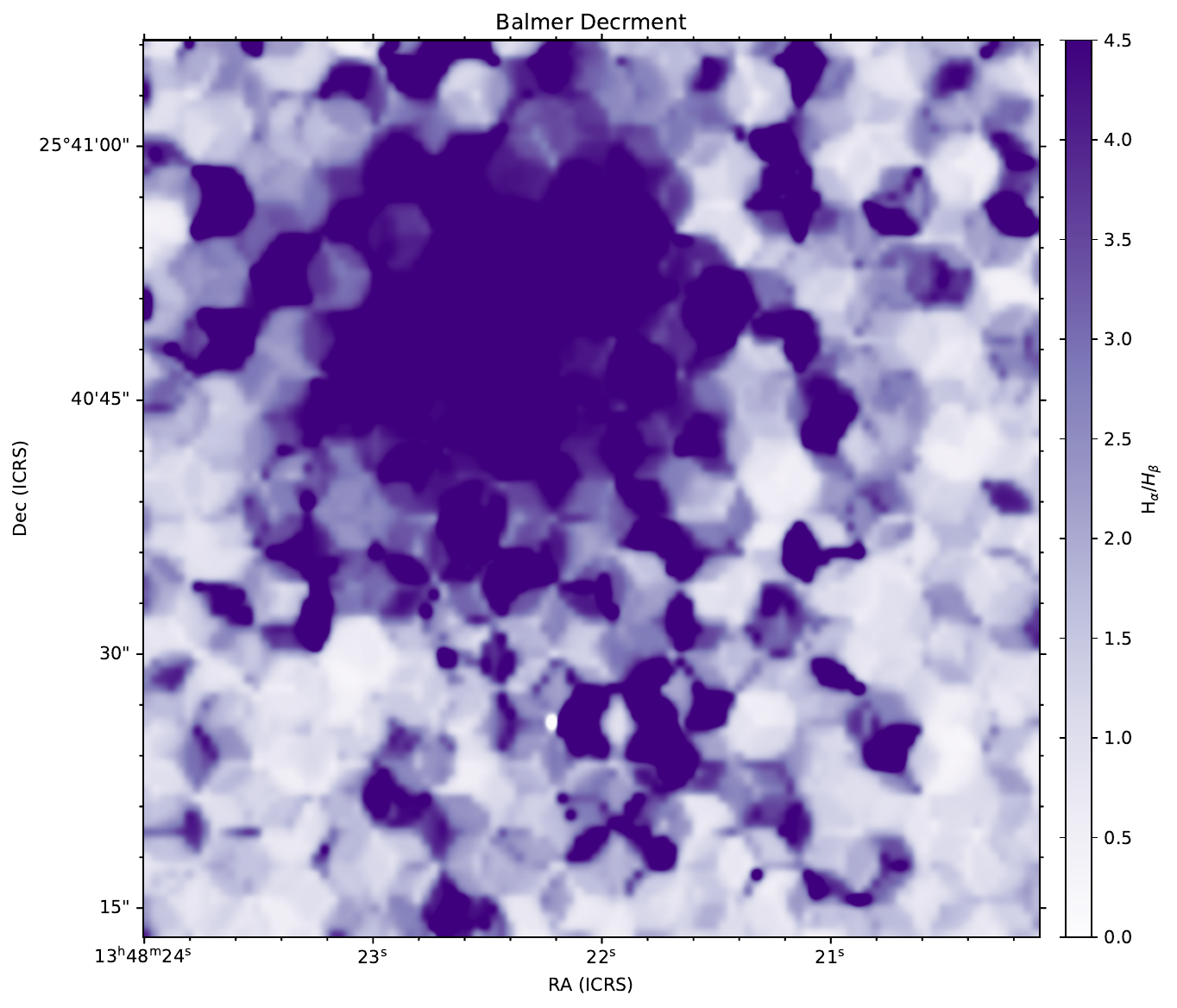}
    \end{minipage}
    \hfill
    \begin{minipage}{0.49\linewidth}  
        \includegraphics[width=\textwidth, height=\textwidth]{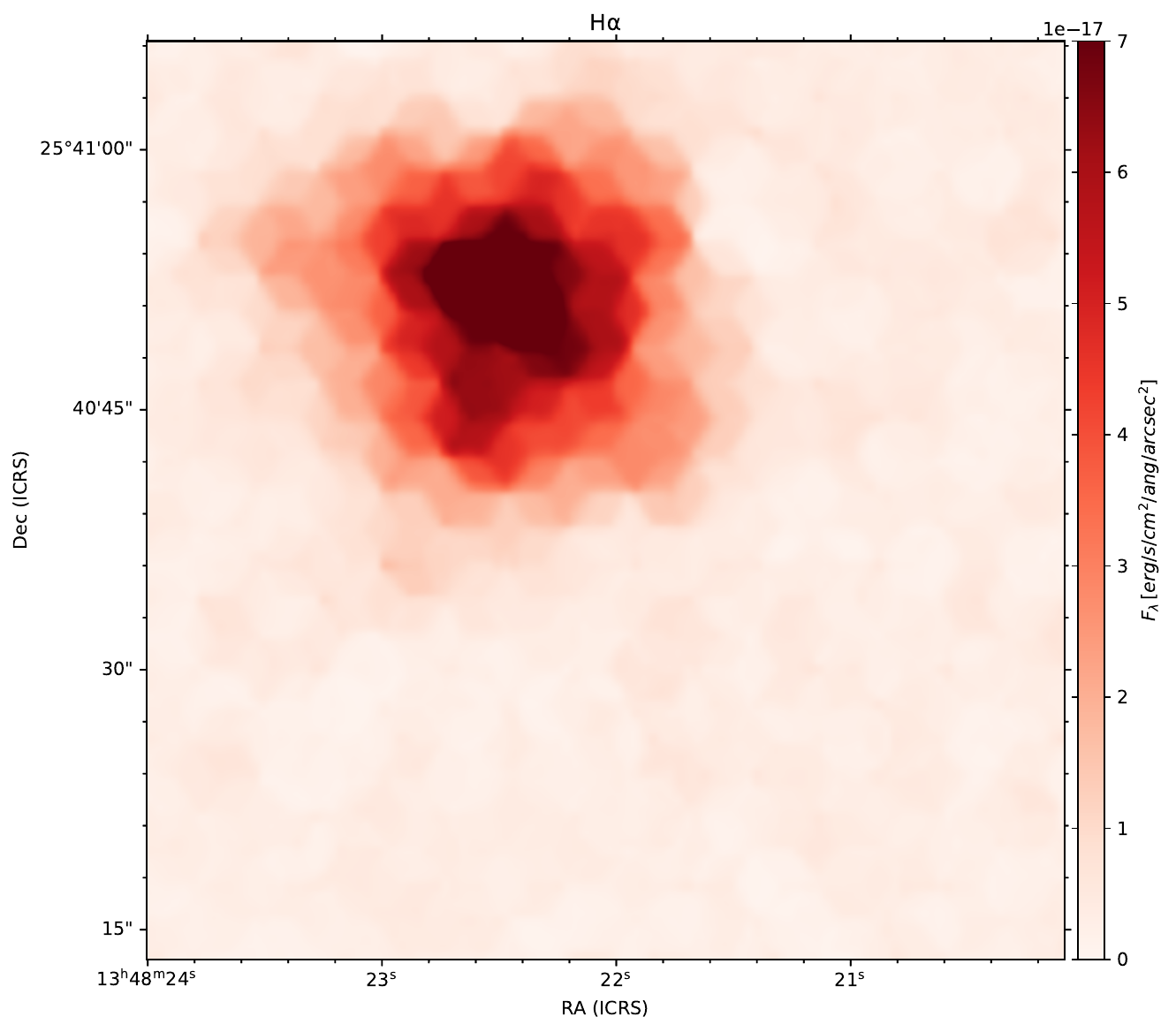}
    \end{minipage}
    \hfill
    \begin{minipage}{0.49\linewidth}  
        \includegraphics[width=\textwidth, height=\textwidth]{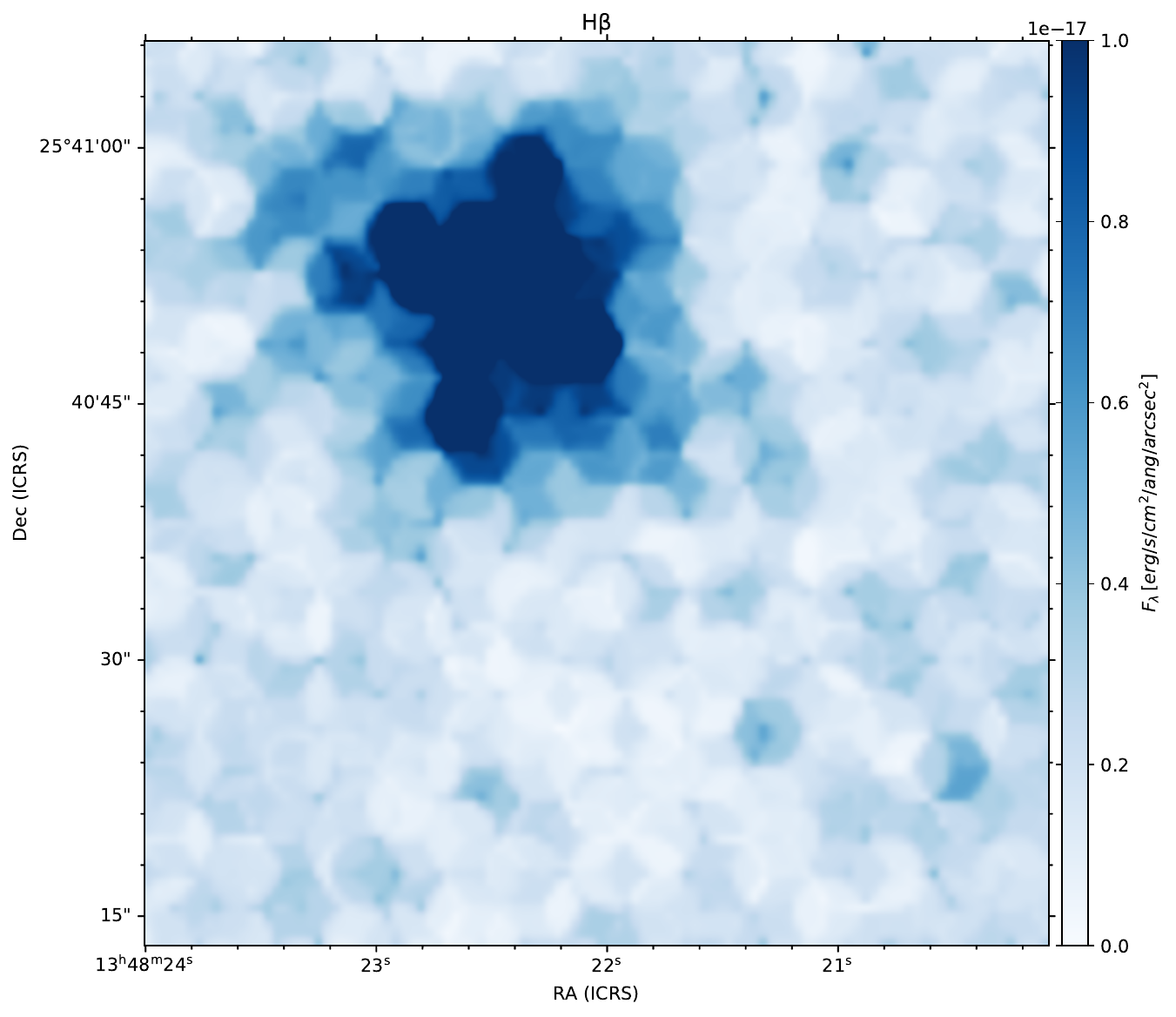}
    \end{minipage}
    \hfill
    \begin{minipage}{0.49\linewidth}  
        \includegraphics[width=\textwidth]{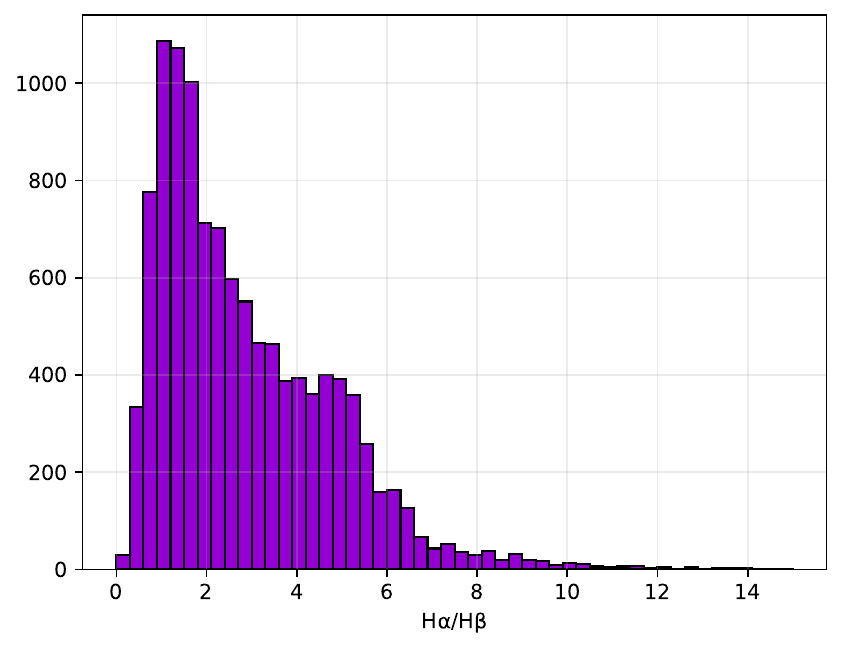}
    \end{minipage}
    \begin{minipage}{\linewidth}  
    \begin{center}
      \caption{ Top left: The Balmer Decrement (H$\alpha$/H$\beta$) map of VV\,191b. The ratio values are ranged from 0.5-4 (the theoretical value for no attenuation is 2.86). The following two maps are the \Ha\ map (top right) and the \Hb\ map (bottom left) of the spiral foreground. Bottom right: histogram of over 11,000 values for the Balmer decrement values over the same regions as the three maps, ranged from 0-14. The hexagonal shapes are the GCMS fibers projected on the spaxel map. }
      \label{fig:3 maps}
    \end{center}
    \end{minipage}
\end{figure*}

Figure \ref{fig:3 maps} shows the three maps generated from the GCMS spectra smoothed with a Gaussian interpolation. We can easily see the spiral foreground galaxy as it is present in all three maps. The elliptical background galaxy, VV\,191a, isn't seen due to its lack of Balmer line emission. A total of 11,236 spaxels GCMS IFU observations are processed in Figure \ref{fig:3 maps}, giving the same number of Balmer decrement values to map. The histogram of the Balmer decrements is the bottom right of Figure \ref{fig:3 maps}. The majority of the small values (\BD\ $\leq$ 0.5) comes from the dark night sky surrounding the galaxy pair. 

The maps differ the most in the individual \Ha\ and \Hb\ maps. \rev{Fortunately,} the center of VV\,191b has the darkest regions with a consistent and uniform fade in color radially outward. \rev{We expect the \BD\ distribution to increase in scatter as the S/N decreases, so it should follow that as the galactocentric radius increases, the \BD\ scatter should increase as well, as the noisier spaxels should reside in the outer spiral arms. The observed correlation confirms that the method of emission line extraction gives accurate results.} Visually, a cleaner distribution surrounds the spiral galaxy from the \Ha\ map, indicating a major source of a noisy background derives from \Hb\ measurements. A noisy \Hb\ map is not unexpected as there is much more absorption surrounding \Hb. Figure \ref{fig: spectra} shows the \Hb\ absorption surrounding its peak that hides its relative flux \citep{groves_balmer_2012} as compared to the \Ha\ emission of that same spaxel.


\begin{figure}[h!]
    \begin{minipage}{.5\textwidth}
            \includegraphics[width=\textwidth, height=.75\linewidth]{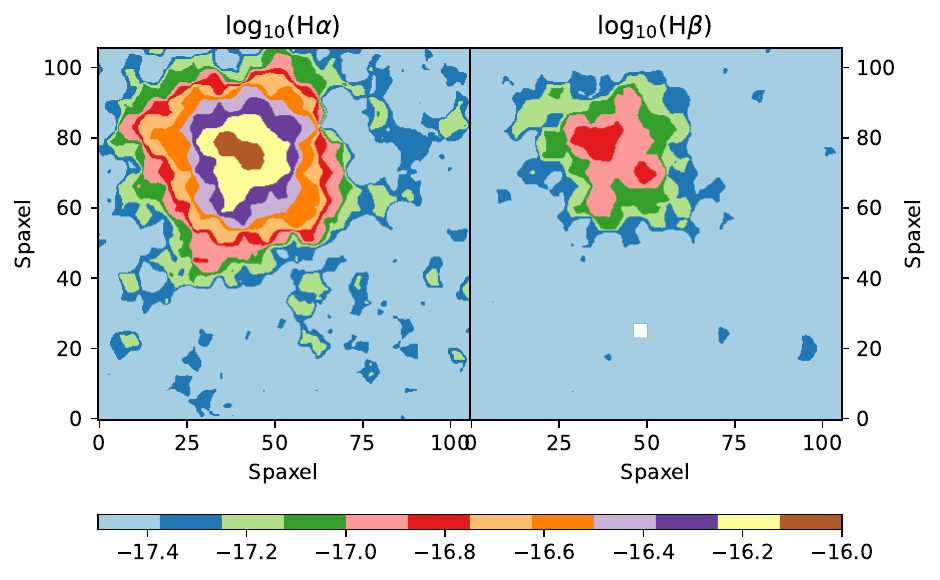}
    \end{minipage}
    \begin{minipage}{.5\textwidth}
            \caption{The logarithm base 10 of the \Ha\ and \Hb\ maps of the top right and bottom left panels of Figure \ref{fig:3 maps}. The maps cover the same area as Figure \ref{fig:3 maps} but are now labeled in pixels instead of WCS coordinates to give perspective on size of pixels. Both maps are smoothed using the Gaussian interpolation \rev{(3 pixels across)}.}
            \label{fig:log maps}
    \end{minipage}
\end{figure}

\begin{figure*}
    \begin{minipage}{0.32\textwidth}  
        \includegraphics[width=\textwidth]{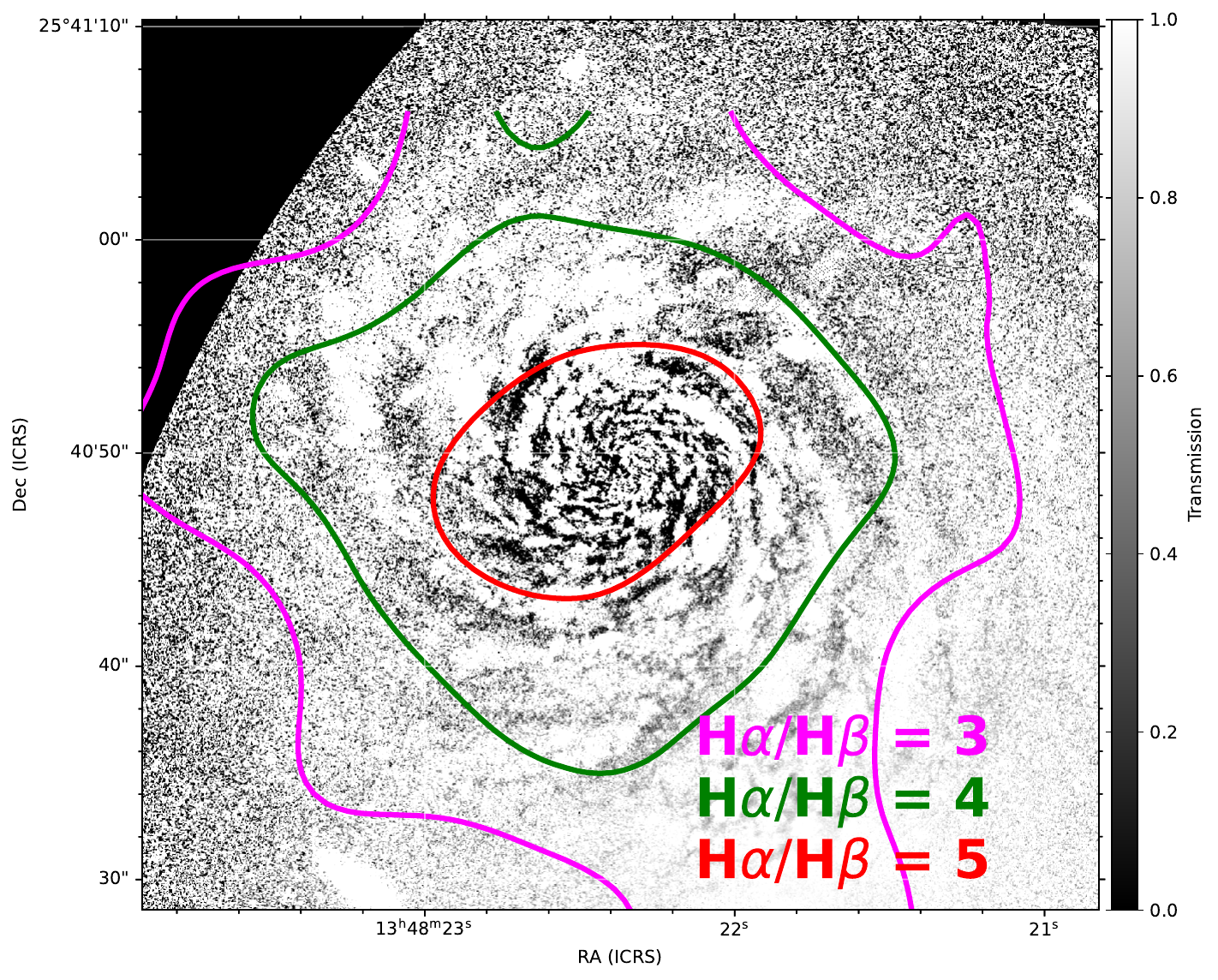}
    \end{minipage}
    \begin{minipage}{0.32\textwidth}  
        \includegraphics[width=\textwidth]{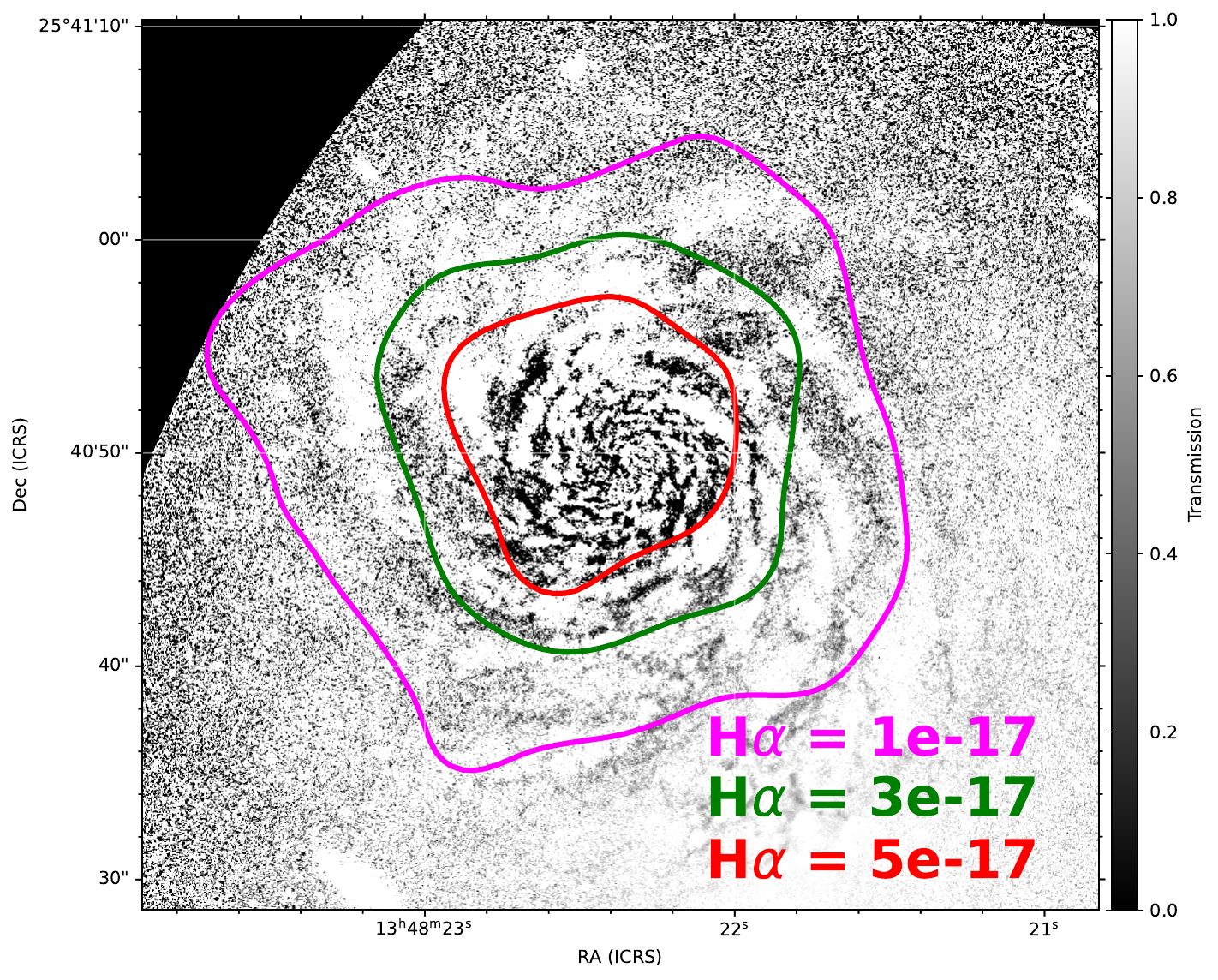}
    \end{minipage}
    \begin{minipage}{0.32\textwidth}  
        \includegraphics[width=\textwidth]{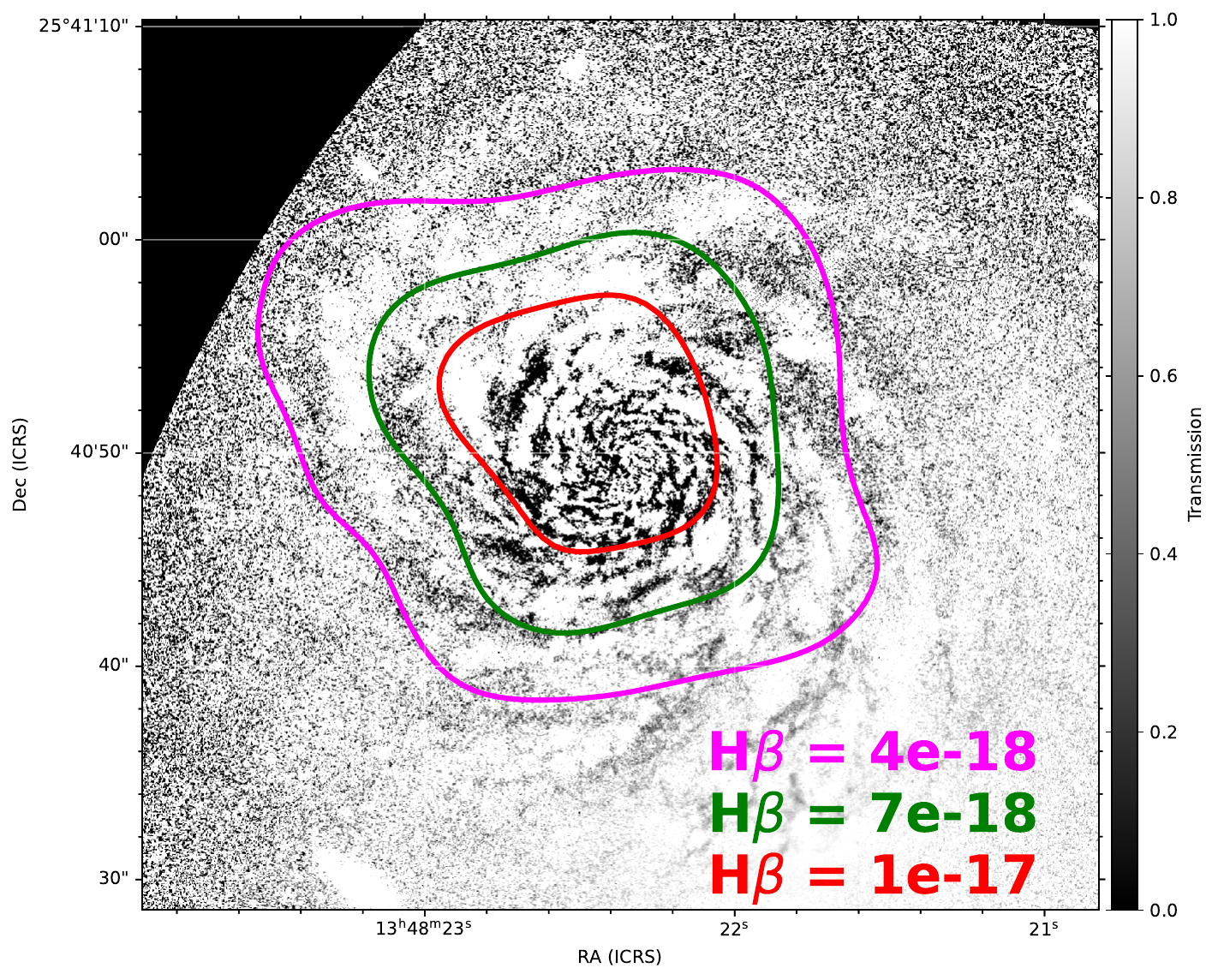}
    \end{minipage}
        \caption{The three panels show the dust map of VV\,191b spiral foreground galaxy in the F606W filter from HST with different contours from the Balmer emission lines: the ratio (\BD, left panel), \Ha\ (middle panel), and \Hb\ (right panel) individually, respectively. Color-bars display the transmission from the grayscale dust maps, ranging from 0-1, with 1 meaning full transmission (no dust). Left: Balmer Decrement contour map with three different levels: The {\color{magenta} magenta} level represents H$\alpha$/H$\beta$ = 2.75, {\color{green} green} represents H$\alpha$/H$\beta$ = 3, and {\color{red} red} has H$\alpha$/H$\beta$ = 3.75. The H$\alpha$ (middle) and the H$\beta$ (right) contour maps are the other two figures. Note that the latter two maps are on the scale of 10$^{-17}$ (\Ha) and 10$^{-17}$ (\Hb) with levels labeled on the figures. \rev{The maps of the separate Balmer lines by themselves have small peak detection values after subtracting out the stellar continuum. $H\alpha$ and $H\beta$ lie in the range of $\sim 10^{-18} - 10^{-17} erg s^{-1} cm^{-2}$ \AA$^{-1}$ arcsec$^{-2}$ for the spiral galaxy, respectively. For the \Ha\ contour, the three levels are {\color{magenta} $1 \times 10^{-17}$ in  magenta}, {\color{green} $3 \times 10^{-17}$ in  green}, and {\color{red} $5 \times 10^{-17}$ in  red}. Similarly, for the \Hb\ contour, the three levels displayed are {\color{magenta} $10^{-18}$}, {\color{green} $4^{-18}$}, and {\color{red} $7 \times 10^{-18}$}.}}
        \label{fig:contours}
\end{figure*}



\begin{figure}[ht]
    \centering
    \includegraphics[width=.5\textwidth]{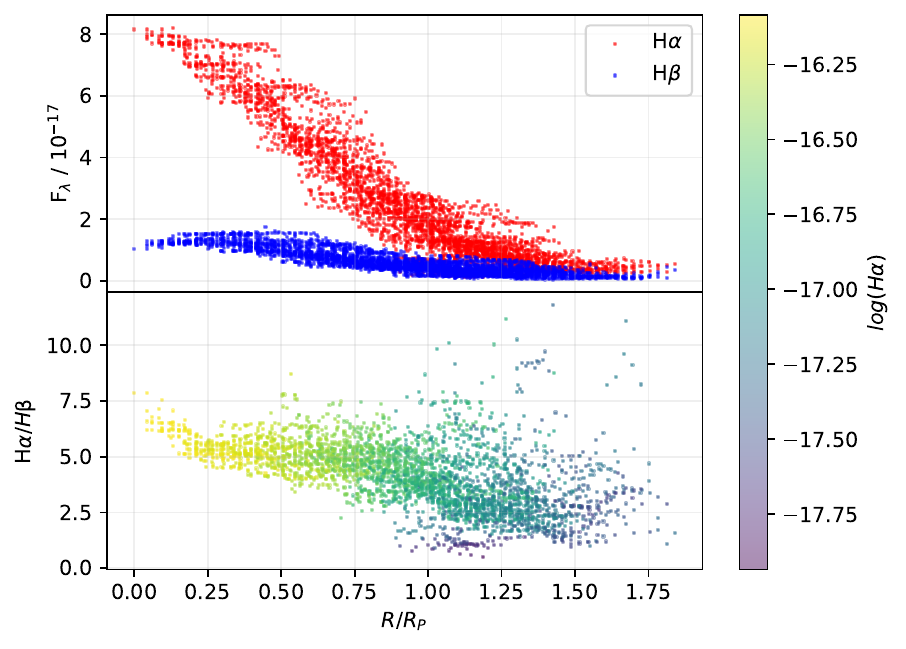}
    \caption{Top: Flux for each individual emission line for both \Ha\ (red) and \Hb\ (blue) for the entire foreground galaxy VV\,191b. These emission line measurement are constrained all the way to the center spaxel (R/R$_{P}$ = 0). Units for flux (F$_\lambda$) are $erg s^{-1} cm^{-2}$ \AA$^{-1}$ arcsec$^{-2}$ and divided by a normalization factor of 10$^{-17}$.  Bottom: Balmer decrement values for each spaxel in the dust lane region are matched with the distance to the center spaxel of VV\,191b spiral foreground galaxy and then converted to distance in kpc by the WCS.  }
    \label{fig:bd vs radius}
\end{figure}

We plot the values in the maps displayed in Figure \ref{fig:3 maps} as overlaid contours in Figure \ref{fig:contours}. The contours overlay the F606W dust map provided by \cite{keel_jwsts_2023}. 
Each contour map has three levels, which differentiate from each other due to the differences in detection values. The levels are indicated in the figure and caption. Figure \ref{fig:bd vs radius} gives all of the \BD\ values (bottom) along with plotting every single emission line (top) for VV\,191b. With variation throughout, the Balmer decrement steadily decreases as the radius from the center spaxel of the spiral galaxy increases. The radius is normalized by the Petrosian radius: $R_{P} = 12$ kpc \citep{keel_jwsts_2023}.

\subsection{Measurements of the Galaxy Scale-Length of VV 191b}
\label{s:scale length}

Jos\'{e} Luis S\'{e}rsic's R$^{1/n}$ model \citep{sersic_influence_1963, sersic_atlas_1968} has been used to profile galaxies for decades. It remains an effective method to learn and gain detailed information on galaxy morphology \citep{blanton_estimating_2003}. For this paper, we follow the approach from \cite{graham_concise_2005}. We express the $R^{1/n}$ model in terms of the surface brightness profile, $\mu(R)$, as a function of the radius from the galaxy center, R: 

\begin{equation}
    \mu(R) = \mu_{0} + \frac{2.5}{\ln{10}} \left (\frac{R}{h} \right)^{1/n}
    \label{eq mu scale}
\end{equation}

$\mu_{0}$ is the central surface brightness, n is the S\'{e}rsic index, and h is the galaxy scale-length. VV\,191b is a face-on spiral disk, which qualifies as n = 1 S\'{e}rsic profile, the exponential profile. This simplifies Equation \ref{eq mu scale} to just:

\begin{equation}
    \mu(R) = \mu_{0} + 1.086\left (\frac{R}{h} \right)
    \label{eqmu}
\end{equation}

The scale length of a galaxy disk is an effective parameter in galaxy morphology in general. It has been analyzed and calculated in many ways \citep{fathi_scalelength_2010, koopmann_comparison_2006}. Our purpose is to derive the scale length through the surface brightness by using a few independent methods: the Balmer decrement, the \Ha\ emission line, a flat stellar continuum in the V-band ($\lambda_{rest} = 5500$ - $6000$ \AA), and A$_{V}$ from HST photometry in the F606W filter. 

The apparent magnitude is a measure of flux, $\mu$, in units of magnitude arcsec$^{-2}$, is derived for each quantity besides $A_{V}$. $A_{V}$, already related to the logarithm of the flux (Equations \ref{eq transmission} and \ref{eq av and bd}), cannot be derived similarly to the other three. Attenuation is the extinction and scattering of light; therefore, we expect it to have a negative slope as compared to Equation \ref{eqmu} to the surface brightness since $A_{V}$ directly counters the effects of luminous intensities exhibited by the host galaxy. The scatter in the attenuation is significant outside of $1.25R_{P}$, linear regression is only plausible when using the data within this constraint (see Appendix for regression plots). Results and uncertainties for each measurement of the scale length are calculated from the linear regression residuals. They are shown in Table \ref{tbl-scale lengths}.  
\begin{deluxetable}{lccccc}[h]
\tablecaption{Scale-length values of VV\,191b in kpc}
\label{tbl-scale lengths}
\setlength{\linewidth}{7pt}
\setlength{\tabcolsep}{5pt}
\tablehead{
\colhead{h (\BD)} & \colhead{h (\Ha)} & \colhead{h (V)} & \colhead{h (A$_{V}$)} \\ 
\colhead{(kpc)} & \colhead{(kpc)} & \colhead{(kpc)} & \colhead{(kpc)}\\}
\startdata
\rev{$26.2 \pm 6.39$} & \rev{$5.27 \pm 32.3$} & $4.73 \pm 16.0$ & $38.2 \pm 361$  \\
\enddata
\end{deluxetable}


\subsection{Large Dust Lane Region in VV 191b}

VV 191 is a rare case of an occulting galaxy pair with partially overlapping and non-overlapping regions. One interesting physical attribute is that the foreground galaxy, VV\,191b, is a rotationally-symmetric spiral galaxy with clearly visible dust lanes with enough partial overlap and non-overlap regions. This characteristic, analyzed by \cite{keel_jwsts_2023}, presents an opportunity to take a detailed look at regions in the outer arms. We selected a specific sample region from the population of dusty outer arms where the dust lane is backlit by the background elliptical and prevalent while not too close to the center of VV\,191b so that the dust lanes are distinct. 

\rev{Our hand-picked data is specifically selected in the overlapping region so the background elliptical galaxy, VV\,191a, provides the proper back-light to measure the light lost in the line of sight for SED fitting. This smooth, background source differs from using the surrounding stellar continuum in VV\,191b as the light weight. Assuming there is no interaction between the two galaxies (see Section \ref{s:grav interact}), the dust within the spiral arms of VV\,191b is completely independent from its background light source. This independence of source lighting makes the unique arrangement of galaxies a prime candidate to compare the Balmer decrement, used to calculate dust attenuation in dusty HII regions, and dust attenuation from SED fits and the reddening of the stellar continuum \citep[computed for VV\,191b by][]{keel_jwsts_2023}. }   \\

\begin{figure*}[h!]
\begin{minipage}{.48\linewidth}
    \includegraphics[width=\textwidth]{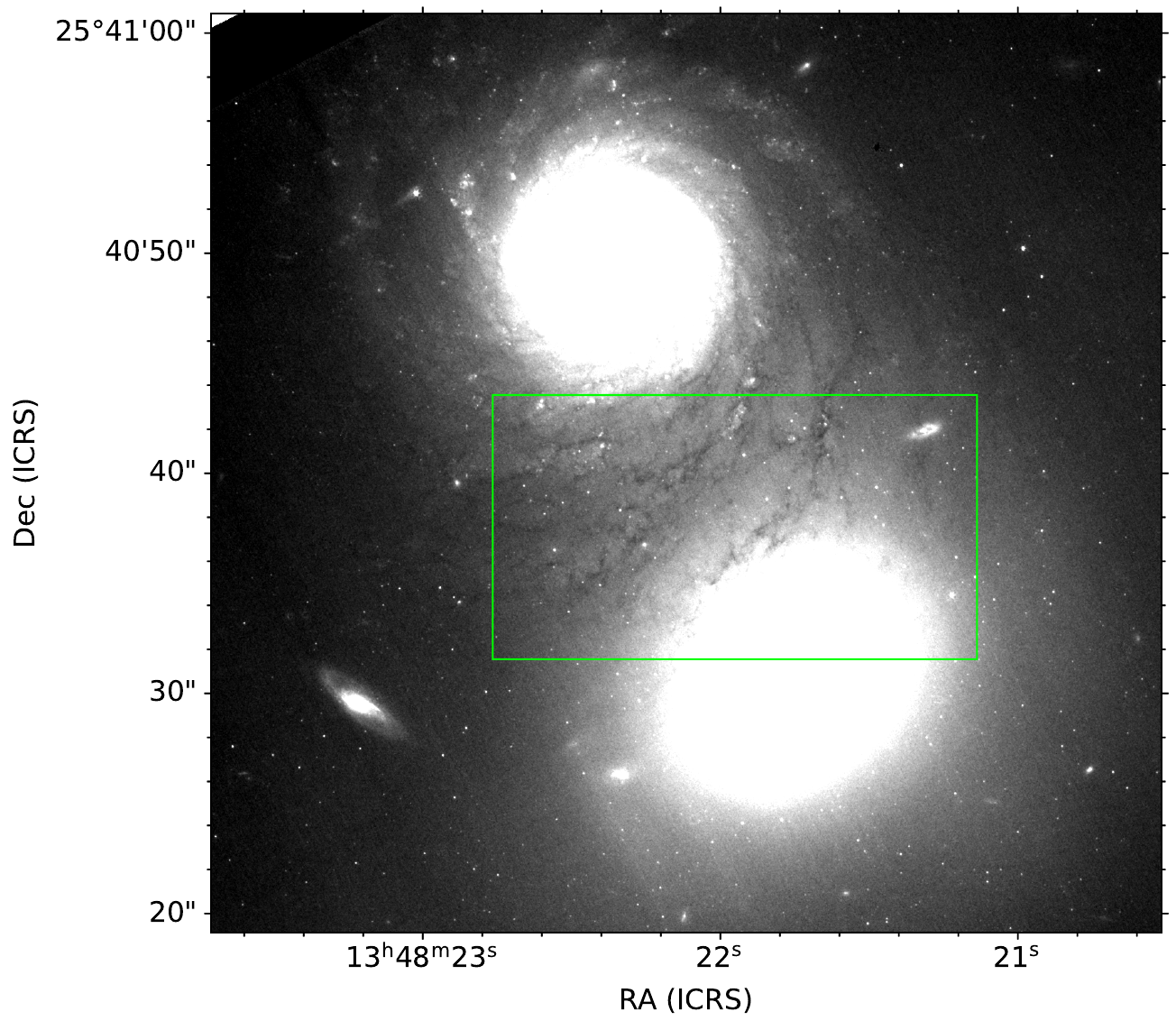}
\end{minipage}
\hfill
\begin{minipage}{.48\linewidth}  
    \includegraphics[width=\textwidth]{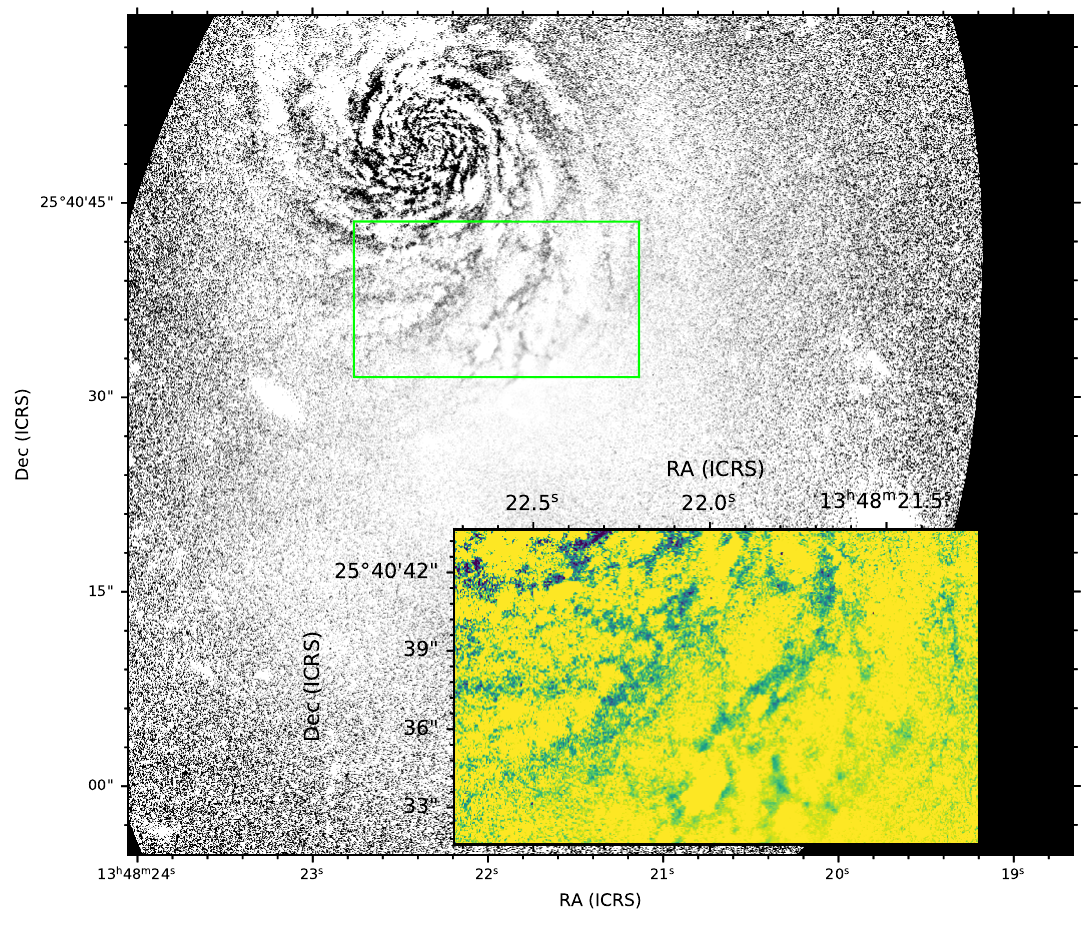} 
\end{minipage}
\hfill
\begin{minipage}[rt]{\textwidth}
    \caption{Left: JWST F090W image of the VV\,191 occulting pair that encloses the area in which we selected our dust lane region. Right: JWST F606W dust map with the region zoomed in the right corner.} 
    \label{fig:DL box from JWST}  
\end{minipage}
\end{figure*}


\subsubsection{Dust Attenuation with the Balmer Decrement in Dust Lane Region}
\label{s:dl}

\begin{figure*}
    \begin{minipage}{.48\textwidth}
        \includegraphics[width=\textwidth]{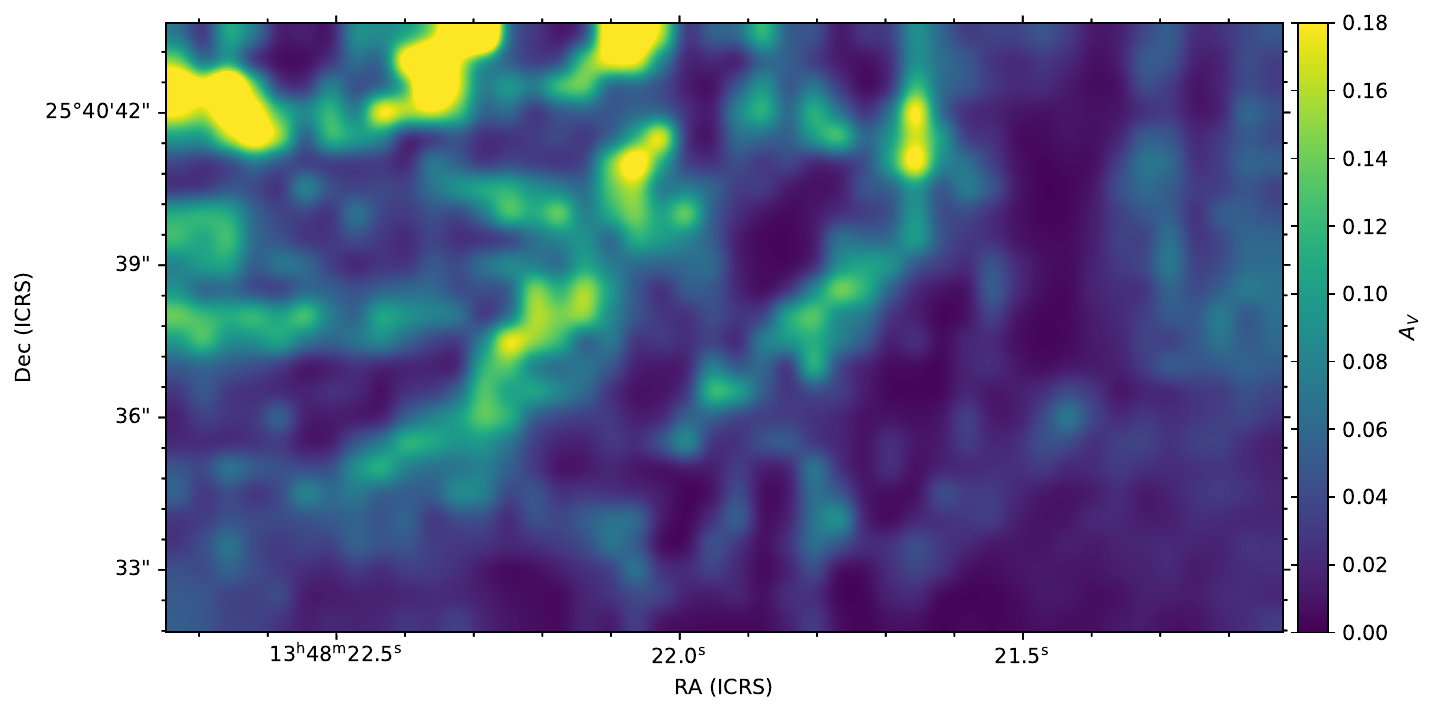}
    \end{minipage}
    \hfill
    \begin{minipage}{.48\textwidth}
        \includegraphics[width=\textwidth]{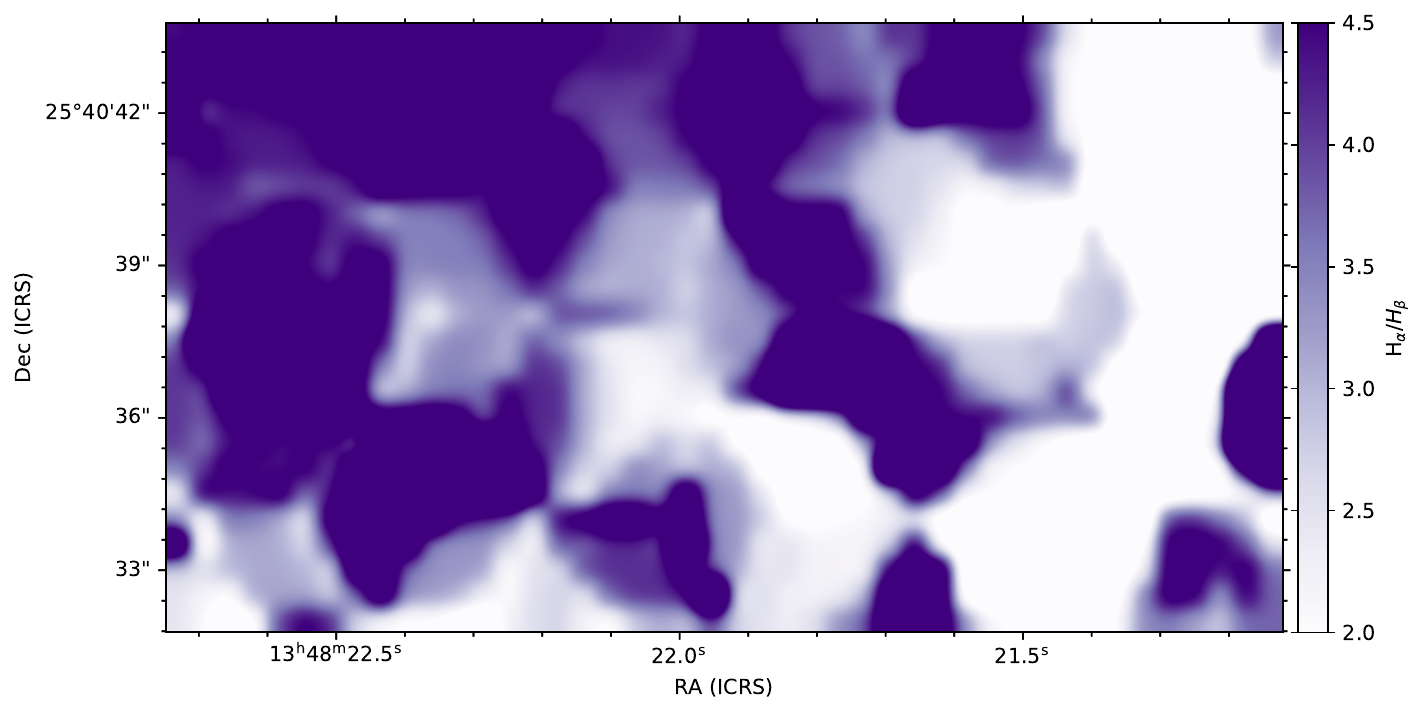}
    \end{minipage}
    \hfill
    \begin{minipage}{\textwidth}
          \caption{The dust lane region highlighted in Figure \ref{fig:DL box from JWST} is shown in detail. Left: The A$_{V}$ map derived from HST/JWST imaging is shown. Right: map of Balmer decrement in dust lane region ranging values from 2-4. The left figure is smoothed using Gaussian interpolation. The attenuation is derived from the pixelated dust lane region, meaning the size of spaxels between the left and right figures are the same.}
    \label{fig:pix_t and BD of DL}
    \end{minipage}
\end{figure*}

Figure \ref{fig:DL box from JWST} shows the hand-picked dust region for further analysis. We chose to select this region to cover a majority portion of the overlap between the two galaxies that is zoomed in enough to effectively study individual dust lanes within VV\,191b. Both images come from HST and JWST. The left image is from the F090W filter and has a wavelength of around 0.9-1 micron. The right image shows the dust map derived from the HST image in the F606W filter (around 0.6 microns). We extrapolated this dust lane region from the F606W dust map through SAOImage DS9. Figure \ref{fig:pix_t and BD of DL} shows the region zoomed-in. To compare the independently derived Balmer decrement and attenuation, the data must match. However, an initial problem presented itself: the spaxels from GCMS are much larger than HST spaxels. \rev{One must match the spatial resolution of the two datsets by finding an accurate relation between the two instruments' spatial resolution. We find that one spaxel from GCMS $\simeq$ 16 $\frac{2}{3}$ HST spaxels. We therefore bin HST spaxels into $16 \frac{2}{3} \times 16 \frac{2}{3}$, rounded to the nearest integer, square sub-matrices. The sigma clipped mean of $\sigma = 10$ of each box takes the transmission value of the new HST 'super-spaxel.'}


\begin{figure}
    \begin{minipage}{.5\textwidth}
        \includegraphics[width=\textwidth]{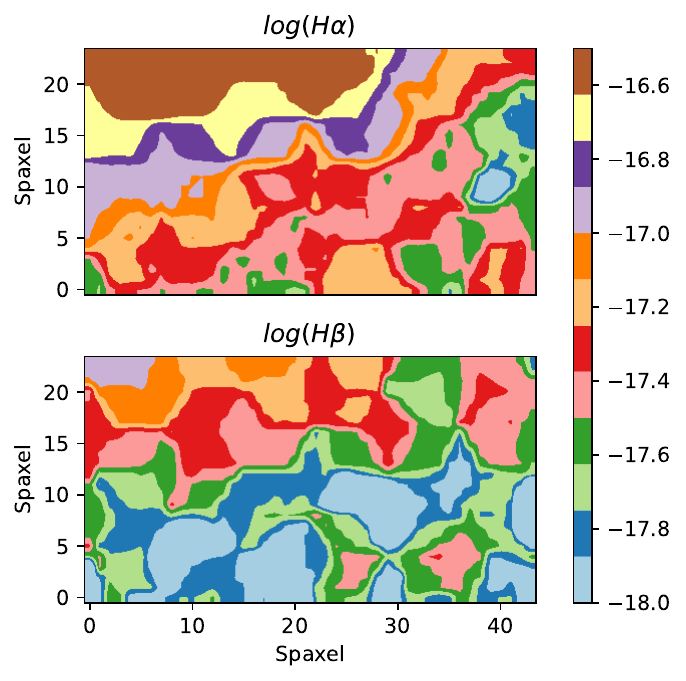}
    \end{minipage}
    \begin{minipage}{.5\textwidth}
        \caption{\Ha\ and \Hb\ maps of the dust lane region in spaxels, respectively. Similar to Figure \ref{fig:log maps}, these maps are shown in spaxels to show perspective on spaxel size as well as the size of the dust region to compare to the WCS coordinates in Figures \ref{fig:DL box from JWST} and \ref{fig:pix_t and BD of DL}. This figure displays the zoomed in dust region, while Figure \ref{fig:log maps} shows the full VV\,191 galaxy pair. The white space in the two figures indicate emission line readings of $< 10^{-18} $. The maps are given in log units with a Gaussian interpolation \rev{(3 pixels across)}.}
        \label{fig:DL stacked log maps}        
    \end{minipage}
\end{figure}     

\begin{figure}
    \includegraphics[width=.49\textwidth]{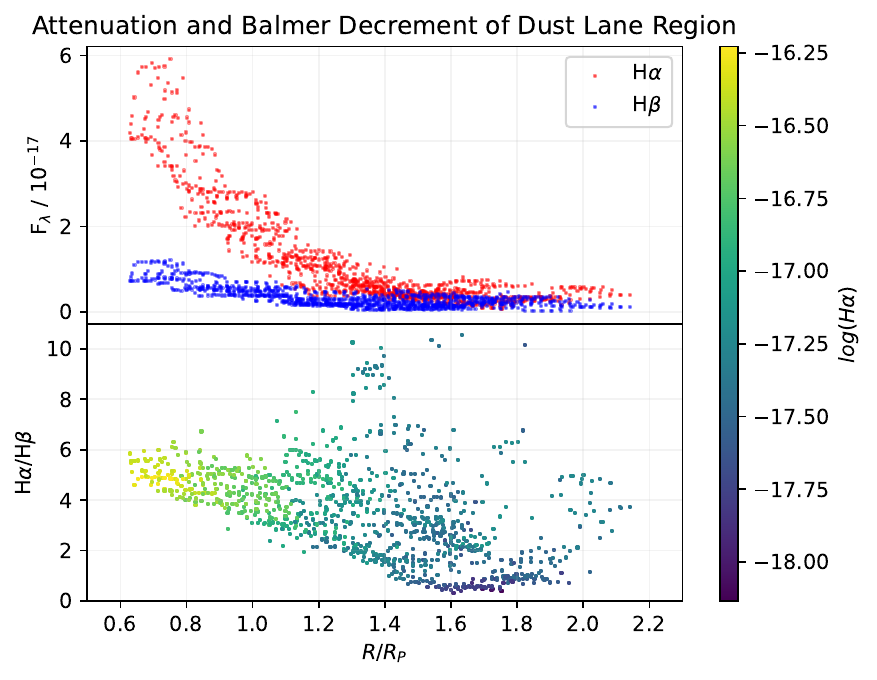}
    \caption{Data gathered from the hand picked region shown in Figure \ref{fig:DL box from JWST}. The Balmer line fluxes (top) and the Balmer decrement values (bottom) plotted against the radius from the center spaxel of VV\,191b divided by the Petrosian radius ($R_{P} = 12$ kpc). }
    \label{fig:r vs bd DL}
\end{figure}

Figure \ref{fig:r vs bd DL} shows both the Balmer line fluxes and the Balmer \rev{decrement values plotted against} the distance from the center spaxel normalized by the Petrosian radius of VV\,191b, as defined by \cite{keel_jwsts_2023}. The angular size of 1.008 kpc arcsecond$^{-1}$, given by \cite{keel_jwsts_2023}, is the conversion from pixels to kpc \citep{wright_cosmology_2006}. The better the S/N for each spaxel, the more accurate the measurement will be. Hence, the Balmer decrement data has a $\log{H\alpha}$ feature to give indication of S/N (larger \Ha\ flux implies higher S/N). \Ha\ gives a good indication of S/N due to \Ha\ being an optimal emission line feature for our bandwidth with little surrounding absorption \citep{rosales-ortega_integrated_2012}. 

Results from \cite{keel_jwsts_2023} on VV\,191b provide further important information on the reddening slopes found for this spiral galaxy. 

\begin{figure*}
    \centering
    \includegraphics[width=.8\textwidth]{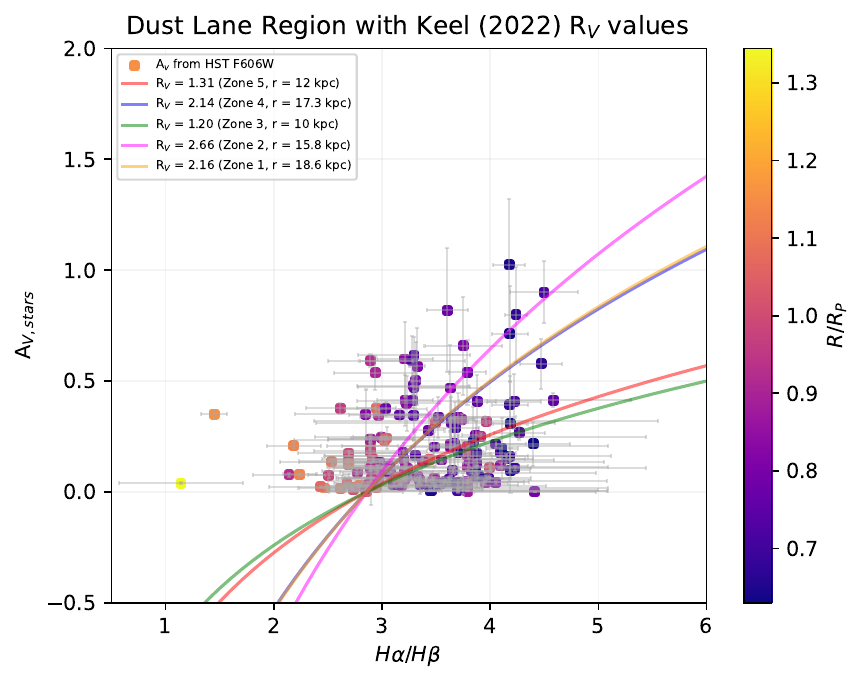}
    \caption{Attenuation, A$_V = -1.086\times \rm{ln(Transmission)}$ derived from HST F606W dust map plot against the Balmer decrement values from GCMS IFU observations. Attenuation relations are plotted using the R$_{V}$ values derived from \protect\cite{keel_jwsts_2023}. Due to matching the number of pixels, and therefore arrays in configuring the data, the ``pixelated'' version of the transmission map is needed.  Zones 1, 2, and 3 are arm regions with organized dust lanes, while zones 4 and 5 are interarm regions with minimal coverage by distinct lanes. Detailed images of the different zones are shown in great detail in \protect\cite{keel_galaxy_2013}, their Figure 8.}
    \label{fig:BOSS1}
\end{figure*}

\begin{figure*}
    \centering
    \includegraphics[width=.8\textwidth]{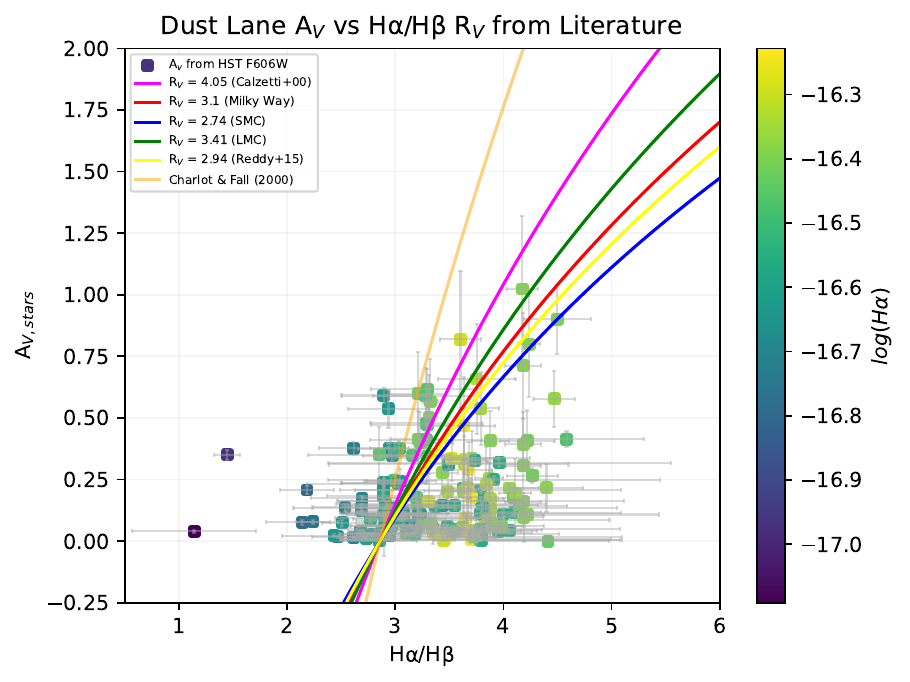}
    \caption{The attenuation (Equation \ref{eq transmission}), derived from HST F606W dust map, plotted against the Balmer decrement values from GCMS. The different reddening slopes, R$_{V}$, from literature. For each slope defined besides \cite{charlot_simple_2000} (relation defined on Equation \ref{eqcf00}), the attenuation curves are calculated using Equations \ref{eq av and bd} and \ref{eq:k values}.}
    \label{fig:BOSS2}
\end{figure*}

Figure \ref{fig:BOSS1} shows A$_{V}$ plotted against \BD\ with comparison to the reddening slopes derived from \cite{keel_jwsts_2023}. Figure \ref{fig:BOSS2} shows the same data with the most commonly used reddening slopes from literature, nicely organized and put together by \cite{nagao_extinction_2016}. \rev{For these figures, we impose a signal-to-noise ratio threshold of $S/N_{H\beta} \geq 2$ for the more uncertain measurement: \Hb. We did not impose the S/N cut in the previous figures to show the galactocentric scatter in our emission line measurements. This cut greatly reduces our sample from 1,056 spaxels to just 167. The cut significantly reduces the \BD\ scatter in low A$_{V}$.} The two figures differ in reddening slopes given and in the color coding. Figure \ref{fig:BOSS1} gives the radius of each spaxel to the center spaxel, while Figure \ref{fig:BOSS2} displays the the \Ha\ emission to give insight on the S/N of each spaxel. Both figures are necessary to show as one figure shows R$_{V}$ results directly for VV\,191b while the other shows the comparison to the commonly used reddening slopes from the literature.


For Figure \ref{fig:BOSS2}, the \cite{calzetti_dust_2000} relation from Equations \ref{eq color excess} and \ref{eq av and bd} are used to find the overarching \BD-dependent attenuation curve for each slope except for \cite{charlot_simple_2000}, for which we use the relation:

\begin{multline} \label{eqcf00} 
A_{V} = -1.086 \times \tau_{V}\\
= -1.086 \times \ln{ \left [\frac{H\alpha/H\beta}{(H\alpha/H\beta)_{int}} \right]}\\
      \times \frac{1}{\left({\lambda_{H\alpha}/5500} \right)^{-.7} - \left({\lambda_{H\beta}/5500} \right)^{-.7}}
\end{multline}

Figures \ref{fig:BOSS1} and \ref{fig:BOSS2} show the same data set, so it is opportune to show different properties (in color coding) when plotting them against the different extinction curves. Figure \ref{fig:BOSS1} shows the distance of each respective spaxel from the center of VV\,191b. The color coding in Figure \ref{fig:BOSS2} shows the strength of the \Ha\ emission line. It is the same color coding as in the bottom of Figure \ref{fig:bd vs radius} to show the S/N strength of each spaxel.



\section{Discussion}
\label{s:discussion}

\rev{\subsection{Offset in Balmer Decrement Zero Point}}
\label{s:hb correction}

\rev{It is well known that the intrinsic value of the Balmer decrement for the parameters mentioned in Section \ref{s:Balmer Dec} is (\BD)$_{int}$ = 2.86 \citep{osterbrock_astrophysics_2006}. Thus, the zero-point for A$_{V}$ should be at this value. However, Figures \ref{fig:BOSS1} and \ref{fig:BOSS2} indicate that this intrinsic value is much higher than 2.86. This offset is interesting, but not too alarming as while our method of lowering the \Ha\ and \Hb\ continuum to extract the emission lines attempted to account for surrounding stellar absorption from the photosphere of A, B, and O stars, it may not have been fully corrected, especially for \Hb. A likely explanation is the absorption that is eaten out from \textit{under} the emission line is not taken into account. This phenomenon is more important for \Hb\ because of its weaker emission \citep{keel_spectroscopic_1983}.}

\rev{An estimation of \Hb\ stellar absorption that is eaten out of the emission line can be made empirically. Emission line fluxes are corrected for underlying absorption effect as follows \citep{kong_spectroscopic_2004}: }

\rev{\begin{equation}
    F_{corr} = F_{obs}(1 + EW_{abs}/EW_{obs})
\end{equation}}

\rev{$F_{corr}$ and  $F_{obs}$ are the respective corrected and observed flux line. $EW_{obs}$ and $EW_{abs}$ are the observed and absorption EWs, respectively. We find that an estimation of $EW_{abs} \sim 1/3 \times EW_{obs}$ accurately shifts the values of \BD\ such that the zero-point of $A_{V}$ implies \BD\ $\sim 2.86$. Figures \ref{fig:BOSS1} and \ref{fig:BOSS2} reflect this empirical shift.}

\rev{\subsection{Reddening Slopes at 1.4 kpc Sampling}}

\rev{In this paper, we evaluate the reddening law (R$_{V}$) with a spatial sampling of 1.4 kpc (see Section \ref{s:data}). This is relatively large to the typical scales of the ISM or HII regions. At larger sampling, the ability to observe multiple {\sc Hii} regions as opposed to individual regions will contribute to averaging out ISM properties of different HII regions. The averaging out of attenuation laws (R$_{V}$) affects measurements of star formation laws, which depend on factors such as Balmer emission lines, set at large physical scales to break down in smaller scales \citep{schruba_scale_2010}. From Figure \ref{fig:BOSS2}, at higher attenuation (A$_{V} \sim 0.5$), \BD\ follows the trends laid out by the provided reddening slopes, given S/N $\geq$ 2. We conclude that the R$_{V}$ slopes provided in Figure \ref{fig:BOSS2} are lower over the 1.4 kpc sampling scale.}

\subsection{Scale Length Fits}

There is significant scatter in the values of the scale length. One of the main reasons for the high error is from the fits themselves. A more robust fit such as a random sample consensus (RANSAC) or Huber regression designed for splitting out outliers supplied the best values for the scale length. However, the estimated error from the fits accounts for all values, including the outliers. It is important to note that, aside from $A_{V}$, a majority of outliers arise outside of 1.25$R_{P}$. Within the Petrosian radius contains fairly consistent data.

The Balmer decrement and the attenuation are much flatter than with \Ha\ and the surface brightness. All four methods in finding the scale length vary outside 1.25$R_{P}$. A more robust linear regression is best for these fits due to the apparent outliers in the outer spiral arms. A$_{V}$ has proven to be the data with the most scatter. Within half of the Petrosian radius contains the most scatter in the A$_{V}$, which is not unexpected: the highest attenuation values should be close to the disk galaxy center. \rev{The similar but} very high values among $h_{H\alpha/H\beta}$ and $h_{A_{V}}$ indicate that, for our data, these two quantities are not ideal for calculation of the accurate scale length, but a good indication of how well these quantities agree with each other.

The hand-picked region effectively encapsulates an area backlit by the background elliptical. VV\,191a does not appear in the emission line maps (Figure \ref{fig:3 maps}), so there is no risk of contamination from the background galaxy. Figure \ref{fig:DL stacked log maps} traces the outer dust arms of the face-on spiral galaxy while specifically \rev{depicting} how well each dust arm is detected and encapsulated by the Balmer lines. Comparisons of Figure \ref{fig:DL stacked log maps} and Figure \ref{fig:pix_t and BD of DL} display how well the emission lines fare with the post-binned HST F606W filter observations. The Balmer lines from GCMS could not detect each dust lane in the spiral arm, but it does distinguish broader areas of higher attenuation. For instance, both images in Figure \ref{fig:pix_t and BD of DL} indicate larger values of the Balmer decrement and higher attenuation above 36 arcseconds in declination and 21.5 arcseconds in right ascension. Figure \ref{fig:DL stacked log maps}, logarithm maps of each Balmer line over the same area, is also in agreement.
The line strength scales are short, similar to the disk's, but the Av and Balmer Decrement both have similar scale lengths (Table \ref{tbl-scale lengths}).

The scale-length of the attenuation and Balmer decrement point to a much flatter profile of dust than typically assumed in the literature \citep[cf.][]{xilouris_are_1999,popescu_modelling_2011}: 1.4 times the scale-length of the stellar disk. The Petrosian radius of VV\,191b is 12 kpc \citep{keel_jwsts_2023}, implying a half-light radius of $\sim$ 6 kpc \citep{graham_concise_2005}. The scale-length of dust measured through both measures imply a ratio of 6, about four times greater than used in SED fits. 

\rev{\subsection{Physical Interpretation}}
\rev{\subsubsection{Balmer Lines vs. Stellar Continuum}}

\begin{figure*}
\begin{minipage}{.48\textwidth}
    \includegraphics[width=\textwidth]{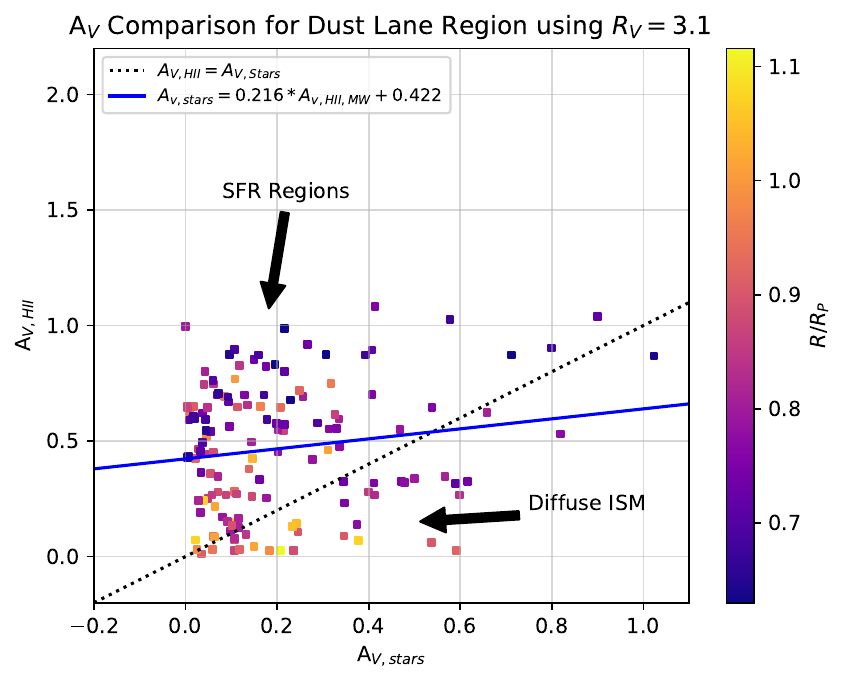}
\end{minipage}
\begin{minipage}{.48\textwidth}  
    \includegraphics[width=\textwidth]{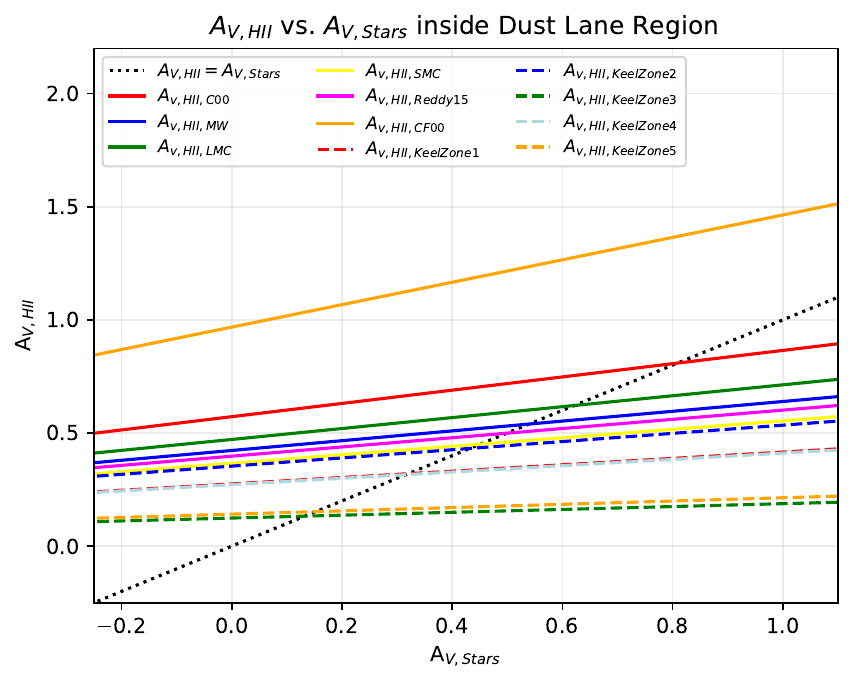}
\end{minipage}
\begin{minipage}{.48\textwidth}
    \includegraphics[width=\textwidth]{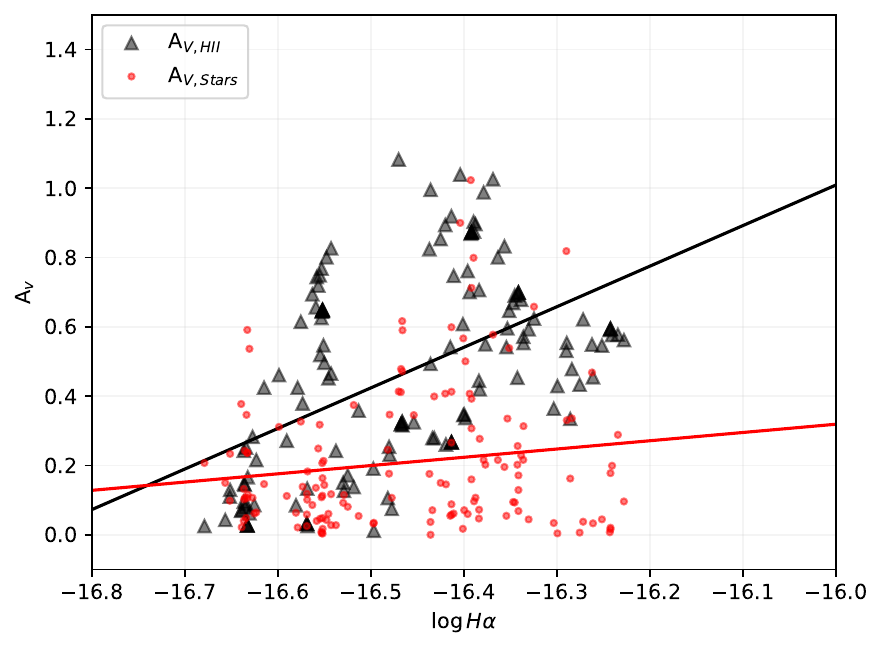}
\end{minipage}
\begin{minipage}{.48\textwidth}
    \includegraphics[width=\textwidth]{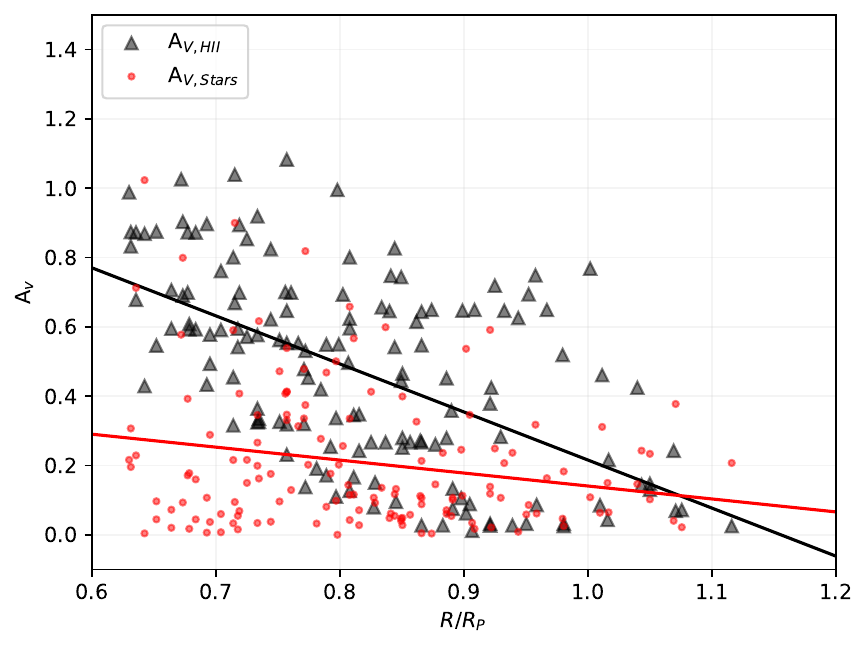}
\end{minipage}
\begin{minipage}{\textwidth}
    \caption{Top left: $A_{V,stars}$, computed from \cite{keel_jwsts_2023}, plotted against $A_{V,HII}$, calculated from Equation \ref{eq av and bd} using the common Milky Way reddening slope $R_{V} = 3.1$ \citep{cardelli_relationship_1989}. The plot is fitted to a linear regression shown by the blue line. The dotted black line represents a 1:1 ratio of attenuations, i.e., $A_{V,stars}$ = $A_{V,HII}$. Top right: Linear regression fits for individual $A_{V,stars}$ vs. $A_{V,HII}$ for each R$_{V}$ value given in Figures \ref{fig:BOSS1} and \ref{fig:BOSS2}. Bottom left: Both attenuations, $A_{V,stars}$ and $A_{V,HII}$ plotted analogous to \cite{kreckel_mapping_2013} Figure 4. We find that the physical scales from \cite{kreckel_mapping_2013} are closest to our own, giving us a direct comparison of our data to other works. Like \cite{kreckel_mapping_2013}, we use R$_{V} = 3.1$. Bottom right: The same data in the bottom left figure but plotted against the radius from the center, normalized by the Petrosian radius.} 
\label{fig:av comps}
\end{minipage}
\end{figure*}

\begin{deluxetable*}{lccccc}
\tablecaption{Ratio of Attenuations from the Literature}
\label{tbl-av_ratios}
\setlength{\tabcolsep}{5pt}
\tablehead{
\colhead{Reference} & \colhead{Survey/Instrument} & \colhead{Sampling} &  \colhead{z} & \colhead{$E(B-V)_{stars}/E(B-V)_{HII}$} \\}
\startdata
\cite{calzetti_dust_1994} & Atlas from \cite{kinney_atlas_1993} & Galaxy-wide &$0.001 \leq z \leq 0.03$ & $0.44$ \\
\cite{yoshikawa_moircs_2010} & MOIRCS & Galaxy-wide & $1 \leq z \leq 2.5$ & $0.26$  \\
\cite{wild_empirical_2011} & SDSS, UKIDSS & Galaxy-wide & $0.067 \leq z_{median} \leq 0.083$ & $0.48$  \\
\cite{kreckel_mapping_2013} & KINGFISH & $\sim 100$ pc & $0.001 \leq z_{median} \leq 0.007$ & $0.47$ \\
\cite{kashino_fmos-cosmos_2013} & FMOS-COSMOS & Galaxy-wide & $1.4 \leq z \leq 1.7$ & $[0.69,0.83]$ \\
\cite{price_direct_2014} & 3D-HST & Galaxy-wide & $1.36 \leq z \leq 1.5$ & $0.53$ \\
\cite{reddy_mosdef_2020} & MOSDEF & Ga;axy-wide & $1.4 \leq z \leq 2.6$ & $0.48$ \\
\enddata
\end{deluxetable*}

\rev{Dust attenuation has been measured through different methods and depictions \citep{salim_dust_2020}. In particular, comparisons between dust attenuation measurements from the reddening of the stellar continuum (A$_{V, stars}$) and attenuation measurements from \BD\ (A$_{V,HII}$) have been investigated throughout the literature (see Table \ref{tbl-av_ratios}). It is important to note that our measurement of A$_{V, stars}$ is unique because the dust in VV\,191b is backlit by VV\,191a, the smooth background elliptical galaxy. Table \ref{tbl-av_ratios} displays notable results between the color excesses derived from the reddening of the stellar continuum ($E(B-V)_{stars}$) and from the Balmer lines ($E(B-V)_{HII}$). We investigate these finding and their interpretations and relate them back to the results to develop our own physical interpretation. }

\rev{Interpretations of the results from the references in Table \ref{tbl-av_ratios} derive from the two-component dust model: one component from the diffuse ISM while the other from short-lived stellar birth clouds. Previous studies (see Table \ref{tbl-av_ratios}) agree that the ratio of color excesses increases with increasing SFR. As SFR increases, the HII regions with higher dust content begin to dominate the ionized gas attenuation, all while the diffuse gas component becomes enriched while dominating A$_{V, stars}$ for both low and high SFR. It is shown by a couple works \citep{wild_empirical_2011,price_direct_2014} that lower specific star formation rate (SSFR) leads to a larger ratio between the color excesses. Low SSFR leads to a larger contribution from older stars that lie outside stellar birth clouds. Since these older stars are not affected by the second component, the stellar continuum will receive less attenuation than nebular emission lines.}

\rev{Most disagreements between A$_{V,HII}$ and A$_{V,stars}$ correlate with higher \Ha\ fluxes that are apparent closer to the center of VV\,191b (bottom right of Figure \ref{fig:av comps}). A$_{V,HII}$ has a steeper galactocentric slope than A$_{V,stars}$, indicating its high dependence on the presence of local O, A, and B stars responsible for larger HII regions of stellar birth clouds. In regions closer to the spiral center that contain higher SFR, we see the two measurements of attenuations differ by a factor of 2, as seen by previous authors. The two attenuations agree with each other in regions where older stars in the stellar continuum are the more dominant feature, contributing to the diffuse ISM component. We see this in the less dusty outer spiral arms further away from the center of VV\,191b. Our measurements agree with other works studying the relationship between the two reddening tracers, though it is highly dependent on the environment within the galaxy itself. Dust attenuation measurements of both the stellar continuum and nebular gas emission at high redshift and large spatial sampling were of the entire galaxies, meaning there was no way to identify separate regions of high and low SFR. Observing entire galaxies tend to favor the dustier, more gaseous centers, which explains the discrepancy of these attenuations for the entire galaxies. Further out, more towards the Petrosian radius, the attenuations become more equal. Stellar light from older stars are the more dominant feature and regions of higher SSFR are more prevalent in the outer spiral arms.} \\

\rev{\subsubsection{Lower Reddening Slopes in the Spiral Arms}}

\rev{The R$_{V}$ slopes contribute to a systematic estimation of A$_{V,HII}$: a lower R$_{V}$ indicate a closer relationship between A$_{V,HII}$ and A$_{V,stars}$ for low values of attenuation (see top right panel of Figure \ref{fig:av comps}). For the case that there is little photon scattering, it is reasonable to assume the prevalence of larger dust grains and fractal cloud structure. The larger grains would simply block out photons while the smaller dust clouds are not big enough to scatter the photons back into the line of sight. In the case that photons are brought back in the line of sight by more scattering would insist the presence of PAHs \citep{nagao_extinction_2016} and Type Ia supernovae (SNe Ia) \citep{goobar_low_2008,conley_is_2007}. We see the agreement in A$_{V,HII}$ and A$_{V,stars}$ for low attenuation in the outer spiral arms near the Petrosian radius (top left and bottom right panels of Figure \ref{fig:av comps}). These outer arms are less likely to contain as many dusty HII regions dominated by young stars to assume multiple scattering. We assume the former argument; larger dust grains and smaller cloud structure unhindered by regions with higher SFR are the more prevalent feature outside the Petrosian radius, leading to lower R$_{V}$ values.}

\subsection{Note on Future Work with Dust Attenuation in VV 191}

To improve the measured relation between Balmer decrement and attenuation, the primary future steps would be to add optical blue filters with the Hubble Space Telescope or IFU to map the Balmer lines in higher spatial resolution. The variance in Balmer decrement and attenuation relation can similarly be explored in a wider range of overlapping pairs, sampling Hubble types and metallicities. The metallicity relation with Balmer decrement and attenuation relation would need improved oxygen line measurements from more sensitive IFU observations.

We plan to expand on the analysis of occulting pair VV\,191 in further follow-up papers to map R$_{V}$ over the smaller practical spatial scales for the HST/JWST observations. Combined imaging of JWST and HST paired with more possibilities of observation time with telescopes or instruments such as MUSE would allow us to further derive attenuation from the occulting galaxy method from \cite{keel_galaxy_2013}. We will use high-resolution imaging to take -- for each pixel in the overlap -- the light lost from the smooth background elliptical galaxy, VV\,191a ($\overline{B}$), which exquisitely highlights dust in the foreground spiral ($F+Be^{-\tau}$) and compare it to its counterpart pixel equidistant from the center on the other side of the galaxy ($\overline{F}$) to derive a separate attenuation slope from the two methods of this paper. The equation used in the derivation, Equation \ref{eq overlap}, is shown here:

\begin{equation}
\label{eq overlap}
\begin{split}  
A_V & = -1.086\times\bar{\tau} \\
    & = -1.086\times\ ln{\left [\frac{(F+Be^{-\tau})-\bar{F}}{\bar{B}} \right]}.
\end{split}
\end{equation}


\rev{In addition, phenomena such as the lower R$_{V}$ in the spiral arms can be explored more with JWST MIRI observations that look into dust composition, specifically dust grain reprocessing within the turbulent ISM. }

In essence, we will aim for more observations on VV\,191 and additional ways to derive an attenuation relation with wavelength for this occulting galaxy pair. This will allow us to probe the changes in attenuation relation to the smallest scales: below 100pc sampling is possible with HST and JWST/NIRCam below 1.5 $\mu$m. 
\\



\section{Conclusions}
\label{s:conclusions}
The attenuation dust curve has many different implications when looking into properties of galaxy morphology, stellar mass, SFR, metallicity, and ISM studies. The observations done by HST F606W and JWST F090W filters combined with referencing and collaborating with PEARLS \citep{windhorst_jwst_2023,keel_jwsts_2023} have given the unique opportunity to provide data analysis of the brand new telescope and its exquisite imaging to the efficient GCMS IFU data with enough spectral coverage to cover the two main Balmer lines to measure \BD. Two different instruments gathered independent data sets to observe and compare two common and useful tools to derive the dust properties of an ideal spiral galaxy VV\,191b. 

Our results are summarized below:

\begin{enumerate}
    \item We present a brief overview of the occulting galaxy pair VV\,191, the relationship between attenuation and Balmer decrement, and the data reduction from GCMS, HST, and JWST. This paper provides the methodology of extracting the Balmer decrement and the multiple uses it provides in analysis of VV\,191b. \\
    \item We present maps and contours of \BD\ along with the \Ha\ and \Hb\ emission lines to show the distribution and strengths of each emission line for the entire foreground spiral galaxy, VV\,191b. \\
    \item Scale lengths were calculated for VV\,191b through the means of \BD , \Ha , the V-band stellar continuum, and A$_{V}$ (Table \ref{tbl-scale lengths}). The large values of the scale length from both the Balmer decrement and attenuation arise from a very flat slope. The large amount of variation in the attenuation very close to the center ($\leq$ 5 kpc) is the most probable reason (See Appendix). The dust scale-length implies a much higher ratio of scales (6x) than previously used in SED fitting. \\
    \item Figures \ref{fig:BOSS1} and \ref{fig:BOSS2} show the main results of comparing the independently derived Balmer decrement and attenuation in the V-band within a large selected region of VV\,191b within the overlap with highlighted dust lanes. \\
    \rev{\item By correcting for the underlying stellar absorption under the \Hb\ emission line by adding in a systematic EW, we correct for the intrinsic Balmer decrement. Our results show evidence on the accuracy for the different R$_{V}$ values given by \cite{keel_jwsts_2023} and from \cite{calzetti_dust_2000, charlot_simple_2000,gordon_quantitative_2003, reddy_mosdef_2015} for physical scales of $\sim$ 1.4 kpc. } \\
    \rev{\item We compute A$_{V, HII}$ from \BD\ measurements using the Milky Way R$_{V}$ to directly compare to A$_{V, stars}$. We find similar results to the literature in that the Balmer decrement is more sensitive to higher attenuation by nearly a factor of 2 near the center of VV\,191b. However, the outer spiral arms (R$\sim$R$_{P}$) show more agreement among the attenuations, albeit at lower values of A$_{V}$. We attribute this finding to larger dust grains and fractal cloud structure in the outer spiral arms.} \\
\end{enumerate}


\section*{Acknowledgements}

\rev{We thank the referee for their helpful comments to improve this paper.}

RAW acknowledges support from NASA JWST Interdisciplinary Scientist grants
NAG5-12460, NNX14AN10G and 80NSSC18K0200 from GSFC.

This research is based in part on observations made with the NASA/ESA Hubble Space Telescope obtained from the Space Telescope Science Institute, which is operated by the Association of Universities for Research in Astronomy, Inc., under NASA contract NAS 5–26555. These observations are associated with program GO-15106 (PI: B.W.~Holwerda).

\rev{This work is supported by NASA Kentucky award No: 80NSSC20M0047 (Graduate Fellowship) awarded to Clayton Robertson and Dr. Benne W. Holwerda.} 

This work is based in part on observations made with the NASA/ESA/CSA James Webb Space Telescope. The data were obtained from the Mikulski Archive for Space Telescopes at the Space Telescope Science Institute, which is operated by the Association of Universities for Research in Astronomy, Inc., under NASA contract NAS 5-03127 for JWST. These observations are associated with program GTO-1176 ``Prime Extragalactic Areas for Reionization and Lensing Science'' (PEARLS, PI: R.~Windhorst)

This work is based in part on observations made with the George Mitchell and Cynthia Spectrograph (GCMS; formerly known as VIRUS-P), hosted at the 2.7 m Harlan J. Smith telescope at the McDonald Observatory \citep{tufts_virus-p_2008, hill_design_2008}.
\rev{Some of the data were obtained from the Mikulski Archive for Space Telescopes (MAST) at the Space Telescope Science Institute. The specific observations analyzed can be accessed via \dataset[DOI]{https://doi.org/10.17909/1h0k-s334}}

This research made use of {\sc NumPy}, {\sc SciPy} \citep{virtanen_scipy_2020}, and {\sc Astropy}, a community-developed core Python package for Astronomy \citep{astropy_collaboration_astropy_2013, astropy_collaboration_astropy_2018, astropy_collaboration_astropy_2022}.









\appendix


\section{More on VV 191b Redshifts}

For the components of VV191 (southern component SDSS J134821.76+254031.0), there are varying redshift values among SDSS data releases. These seem to stem from the combination of light from both galaxies in the fiber nominally pointed at each nucleus and the differences in algorithms for assigning the best redshift value in different data releases. The latest data releases, consistent with inspection of individual spectral features, give $z=0.0513$ for the southern (background) elliptical galaxy (VV191a) and $z=0.0514$ for the foreground spiral (VV191b). The smaller value $z=0.036979$ listed for VV191b by NED (prior to its ingestion of DSS DR13 redshifts) and attributed to \citep{davoust_kinematical_1995} and does not match their published table (their published values match the SDSS data).


This pair may therefore have entered the STARSMOG sample accidentally based on confusion of previous redshift measurements, despite its excellent geometric properties and high level of symmetry. This confusion on the foreground galaxy VV191b's redshift led an incorrect value in the \cite{keel_galaxy_2013} catalog.

We use the new IFU data to re-derive a redshift based on the H$\alpha$ line positions.




\section{Scale length Fits}

\begin{figure}[h]
    \begin{minipage}{0.45\textwidth}  
        \includegraphics[width=\textwidth]{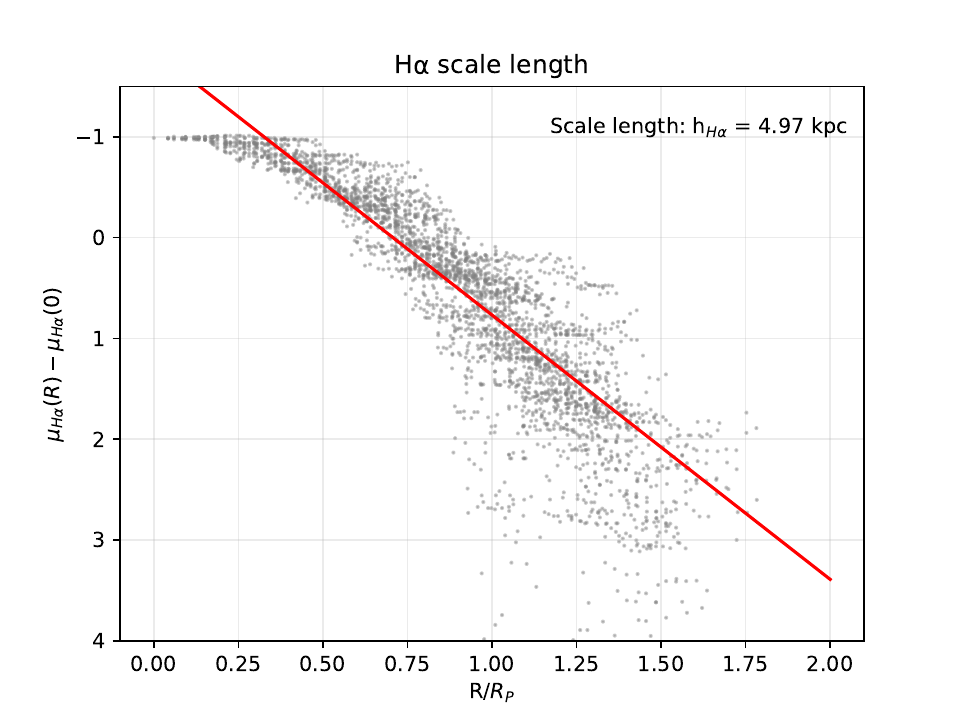}
    \end{minipage}
    \hfill
    \begin{minipage}{0.45\textwidth}  
        \includegraphics[width=\textwidth]{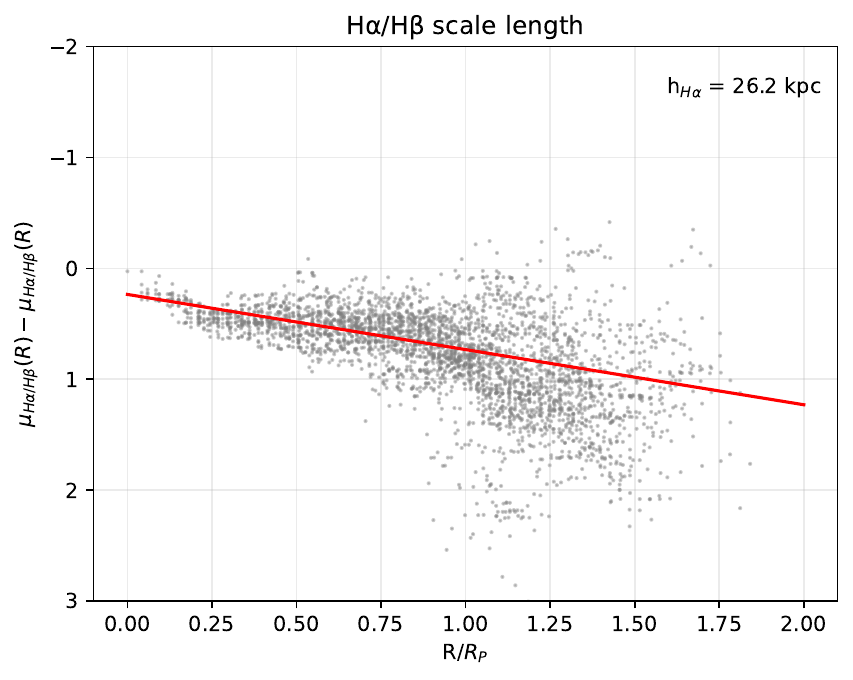}
    \end{minipage}
    \hfill
    \begin{minipage}{0.45\textwidth}  
        \includegraphics[width=\textwidth]{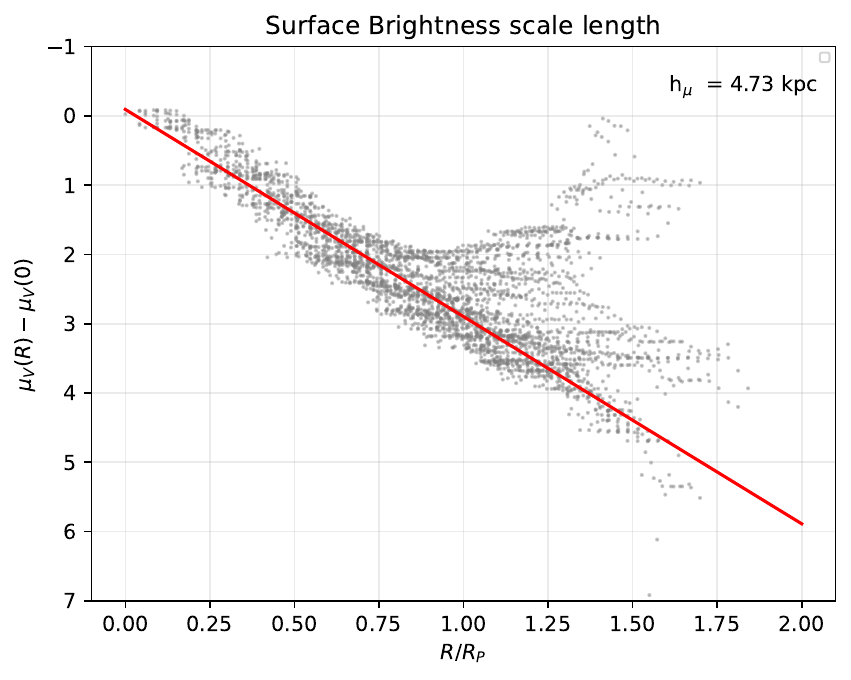}
    \end{minipage}
    \hfill
    \begin{minipage}{0.45\textwidth}  
        \includegraphics[width=\textwidth]{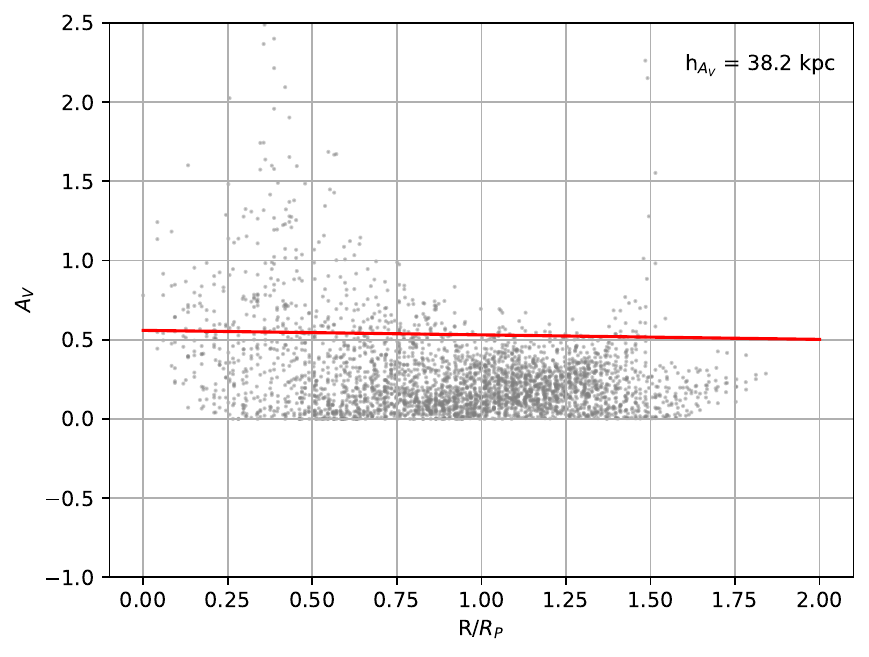}
    \end{minipage}
    \begin{minipage}{\textwidth}  
    \begin{center}
      \caption{Linear fits (red lines) for determining the dust scale length. See Section \ref{s:scale length}. Values for h labeled on each plot are determined from Equation \ref{eqmu}. }
      \label{fig:scale lengths}
    \end{center}
    \end{minipage}
\end{figure}

\newpage









\label{lastpage}
\end{document}